\documentclass[submission, Phys]{SciPost}

\hypersetup{
    colorlinks,
    linkcolor={red!50!black},
    citecolor={blue!50!black},
    urlcolor={blue!80!black}
}

\usepackage[bitstream-charter]{mathdesign}
\DeclareSymbolFont{usualmathcal}{OMS}{cmsy}{m}{n}
\DeclareSymbolFontAlphabet{\mathcal}{usualmathcal}

\usepackage{amsthm}
\usepackage{mathtools}
\usepackage{physics}
\usepackage{xcolor}
\usepackage{bm}
\usepackage{amsmath}
\usepackage{bbm}
\usepackage{array}
\usepackage{graphicx}

\allowdisplaybreaks[1]

\graphicspath{{.}{./Fig/}}

\begin{document}

\begin{center}
{\Large \textbf{Triplet character of 2D-fermion dimers arising
from \textit{s}-wave attraction via spin-orbit coupling and
Zeeman splitting}}
\end{center}

\begin{center}
Ulrich Ebling\textsuperscript{1},
Ulrich Z\"ulicke\textsuperscript{2} and
Joachim Brand\textsuperscript{1$\star$}
\end{center}

\begin{center}
{\bf 1} Dodd-Walls Centre for Photonic and Quantum Technologies,
Centre for Theoretical Chemistry and Physics, New Zealand
Institute for Advanced Study, Massey University,\\ Private Bag
902104, North Shore, Auckland 0745, New Zealand
\\
{\bf 2} Dodd-Walls Centre for Photonic and Quantum Technologies,
School of Chemical and Physical Sciences, Victoria University of
Wellington,\\ PO Box 600, Wellington 6140, New Zealand
\\
${}^\star$ {\small \sf j.brand@massey.ac.nz}
\end{center}

\begin{center}
\today
\end{center}


\section*{Abstract}
{\bf
We theoretically study spin-$\mathbf{1/2}$ fermions confined to
two spatial dimensions and experiencing isotropic short-range
attraction in the presence of both spin-orbit coupling and
Zeeman spin splitting -- a prototypical system for developing
topological superfluidity in the many-body sector. Exact
solutions for two-particle bound states are found to have a
triplet contribution that dominates over the singlet part in an
extended region of parameter space where the combined Zeeman-
and center-of-mass-motion-induced spin-splitting energy is
large. The triplet character of dimers is purest in the regime
of weak \textit{s}-wave interaction strength.
Center-of-mass momentum is one of the parameters determining the
existence of bound states, which we map out for both two- and
one-dimensional types of spin-orbit coupling. Distinctive
features emerging in the orbital part of the bound-state wave
function, including but not limited to its \textit{p}-wave
character, provide observable signatures of unconventional
pairing.
}

\vspace{10pt}
\noindent\rule{\textwidth}{1pt}
\tableofcontents\thispagestyle{fancy}
\noindent\rule{\textwidth}{1pt}
\vspace{10pt}

\section{Introduction}
\label{sec:1}
Since their early days, ultracold atomic gases have provided an
intriguing avenue for exploring and simulating
condensed-matter-physics phenomena. Artificial gauge
fields~\cite{Lin2011,Dalibard2011,Cheuk2012,Wang2012} constitute
pertinent examples for the great degree of control and ability
to fine-tune parameters that are often fixed in a solid-state
material. While spin-orbit coupling for quasi-free electrons in
materials is fundamentally determined by the band
structure~\cite{Elliott1954,Dresselhaus1954,Dresselhaus1955,
Rashba1959} and can only be manipulated in a limited fashion via
nanostructuring~\cite{Winkler2003}, it has recently become
possible to realize synthetic versions of one-dimensional (1D)
\cite{Lin2011} and two-dimensional (2D) \cite{Wu2016,Huang2016}
types of spin-orbit coupling for ultracold neutral atoms by
means of Raman coupling with lasers. Furthermore, advances in
manipulating and probing quantum gases have enabled the study of
low-dimensional systems such as 2D Fermi
gases~\cite{Martiyanov2010,Dyke2011,Ong2015,Ries2015a,
Murthy2015,Mitra2016,Murthy2017,Hueck2018} and provided detailed
insights into their many-body physics via spectroscopic
techniques~\cite{Vale2021a}. In the future, low-dimensional
systems with spin-orbit coupling and Zeeman spin splitting will
allow experimentalists to create exotic condensed-matter phases
such as topological superfluids that can host unconventional
Majorana-fermion excitations \cite{Sato2017}. Mean-field theory
predicts the emergence of such a topological phase in
\textit{s}-wave superfluids of 2D fermions with spin-orbit
coupling and large-enough Zeeman splitting~\cite{Zhang2008,
Sau2010,Alicea2010,Sato2010,Thompson2020}.

Complementary to mean-field studies of interacting many-particle
systems, analysis of the two-fermion bound state in vacuum sheds
a different light on pairing that can provide crucial insight,
e.g., into the strong-coupling (BEC) limit of tightly bound
dimers. Dimers of fermionic atoms are further of interest in
their own right, e.g., as providing qubits for quantum
information processing~\cite{Hartke2021}.  We consider two
particles (atoms) interacting only via isotropic, short range,
and attractive low-energy \textit{s}-wave scattering under the
influence of synthetic spin-orbit coupling. Previous work on
bound states in 3D~\cite{Vyasanakere2011a,Vyasanakere2012,
Yang2019,Dong2013} (see Ref.~\cite{Yin2014} for a trapped
system) and in 2D \cite{He2012,Zhang2012a,Takei2012} largely
ignored the effects of Zeeman spin splitting. Dimers of
spin-orbit-coupled fermions in a 3D gas have already been
produced and probed experimentally~\cite{Fu2014}.

The presence of spin-orbit coupling adds several interesting
features to the two-particle problem, including the fact that
strongly bound states acquire properties solely determined by
the gauge field (becoming so-called
"rashbons"~\cite{Vyasanakere2012}). Furthermore, Galilean
invariance is broken. While total center-of-mass (COM) momentum
is still conserved, it enters the bound-state problem as a
parameter. As a consequence, bound states may dissociate when
scattered to large values of the COM
momentum~\cite{Vyasanakere2012,Takei2012}.  Also, spin-orbit
coupling induces a spin-triplet component in the two-fermion
bound state even with pure \textit{s}-wave
attraction~\cite{Vyasanakere2011a,He2012,Fu2014}. This is in
contrast to the situation in the absence of spin-orbit coupling,
where \textit{s}-wave attraction has no effect in the decoupled
triplet channel and, thus, generates pure spin-singlet bound
states (including in situations with Zeeman splitting present
\cite{Fu2014,Ketterle2008}). As \textit{s}-wave interactions are
usually dominant in ultra-cold atoms where
higher-orbital-momentum interactions are strongly
suppressed~\cite{Pethick2008}, spin-orbit coupling is thus a
promising avenue to induce a triplet character to atomic dimers.

In this paper, we examine the exact bound-state solutions of the
two-particle Schr\"odinger equation in 2D. By considering the
effects of Zeeman spin splitting on the same footing as finite
COM momentum, we extend previous works where the effects of
Zeeman spin splitting were not considered in detail
\cite{He2012,Zhang2012a,Takei2012}. We calculate the dimer
bound-state energy and delineate the critical boundary in
parameter space beyond which Zeeman splitting and/or COM
momentum destabilise the bound state.  We also calculate, for
the first time, the spin projections for the 2D bound states.
The orbital part of the bound-state wave function projected onto
spin-singlet and -triplet components reveals their respective
\textit{s}-wave and \textit{p}-wave-like features. We find that
both finite COM momentum and Zeeman spin-splitting favour
triplet contributions to the ground state and contribute similar
effects to the bound-state problem. However, while finite COM
momentum favours unpolarised triplet character, the Zeeman spin
splitting leads to a spin-polarised triplet character of the
bound state that is associated with a chiral-\textit{p}-wave
orbital wave function. Such triplet-dominated \textit{p}-wave
dimers can be seen as a precursor of the topological superfluid
that is expected to emerge in the many-body
regime~\cite{Zhang2008,Sato2017,Thompson2020}. The region in
parameter space where the triplet character dominates turns out
to be a striking feature of systems with 2D-type (e.g.,
Rashba~\cite{Bychkov1984,Bihlmayer2015,Manchon2015}) spin-orbit
coupling, but triplet-dominated bound states are also present
for 1D-type spin-orbit coupling, albeit in a reduced parameter
range. The polarized spin-triplet character of bound states
could be probed experimentally by spectroscopic techniques
\cite{Fuchs2008,Bergschneider2018} or spin-resolved
momentum-correlation measurements on the single-particle level
in the few-atom regime as recently realised in
Ref.~\cite{Holten2021}.

The remainder of this paper is organized as follows. In
section~\ref{sec:2}, we develop the general formalism for
solving the two-particle problem for any type of spin-orbit
coupling and an effective Zeeman splitting that subsumes both
the actual Zeeman spin splitting and finite COM momentum. In
section~\ref{sec:3}, we apply this formalism to obtain
bound-state solutions for 2D-type spin-orbit coupling. We
first discuss the case with zero COM momentum where we obtain
analytic expressions for all terms in the implicit equation
for the bound-state energy as well as for the critical value of
the Zeeman-spin-splitting energy above which no bound state
exists. We also calculate the relative weights of spin-singlet
and spin-triplet components in the two-fermion bound state and
find that a parameter region exists where bound dimers have a
large triplet component. We then address the case of finite COM
momentum, for which no analytical results can be obtained and
where the shape of the bound state is different. In section
\ref{sec:4}, we repeat our analysis for the case of 1D-type
spin-orbit coupling. Compared with the 2D-type spin-orbit
coupling, the parameter range for having a bound state is
increased. In contrast, while dimers with a dominant triplet
component still exist in this case,
these occur now much closer to the threshold where the combined
Zeeman- and COM-induced spin splitting destabilizes the bound
state. Then, in section~\ref{sec:5}, we plot the orbital part of the
bound-state wave function in relative-momentum space for the
different total-spin components and parameter regimes considered
in the preceding sections. We discuss possible experimental
detection methods in section~\ref{sec:6} before presenting our
conclusions in section~\ref{sec:7}.   

\section{General formalism for solving the two-particle problem}
\label{sec:2}

We consider two spin-$1/2$ fermions that interact via isotropic
short-range interactions. Their movement is confined to the 2D
plane defined by the $x$ and $y$ directions. The particles'
orbital motion is coupled to their spin degree of freedom via a
spin-orbit coupling that depends linearly on in-plane momentum
components. In addition, Zeeman spin splitting lifts the energy
degeneracy of spin projections parallel to the out-of-plane
($z$) direction. The Hamiltonian describing such a two-particle
system is given by
\begin{equation}
\label{eq:2body_hamiltonian}
 \hat{H} = \hat{H}_1^{(1)} \otimes \mathbbm{1} + \mathbbm{1}
 \otimes \hat{H}_1^{(2)} + V(\mathbf{r}_1 - \mathbf{r}_2)\quad ,
\end{equation}
where $V(\mathbf r_1 - \mathbf r_2)$ is the two-particle
interaction potential, and $\hat{H}_1^{(j)}$ denotes the
single-particle Hamiltonian for particle $j$;
\begin{equation}
\label{eq:1body_hamiltonian}
 \hat{H}_1^{(j)} = \frac{\mathbf{p}^2_j}{2m} + h\,
 \hat{\sigma}_z + \hat{\lambda}(\mathbf{p}_j) \quad .
\end{equation}
We use the symbol $\hat \;$ to indicate quantities that are
operators in spin space, such as the vector of Pauli matrices
$\hat{\bm\sigma} \equiv (\hat{\sigma}_x, \hat{\sigma}_y,
\hat{\sigma}_z)$. The Zeeman-spin-splitting parameter $h$
quantifies an energy offset dependent on the $z$ component of
the spin. Spin-orbit
coupling is embodied in the form of $\hat{\lambda}(\mathbf{p})$.
Results obtained in this work pertain to the unitarily
equivalent 2D-Dirac~\cite{Winkler2015},
2D-Rashba~\cite{Bychkov1984,Bihlmayer2015,Manchon2015} and
2D-Dresselhaus~\cite{Dresselhaus1955,Eppenga1988} types of
spin-orbit coupling, as well as the 1D type that is more
straightforwardly realizable in cold-atom
experiments~\cite{Lin2011}. Table~~\ref{tab:soc_types} lists
$\hat{\lambda}(\mathbf{p})$ for each of these four
possibilities.

\begin{table}[t]
\caption{\label{tab:soc_types}%
Types of spin-orbit coupling considered in this work. Each of
these is associated with a particular form of the term
$\hat{\lambda}(\mathbf{p})$ in the single-particle Hamiltonian
(\ref{eq:1body_hamiltonian}) and with a matrix $\mathcal{M}$
entering the transformation of momentum vectors
into momentum-dependent spin splittings
via Eq.~(\ref{eq:qMp}). The constant $\lambda$, which has
dimensions of velocity, quantifies the magnitude of the
spin-orbit coupling.}
\vspace{-0.4cm}
\begin{center}
\begin{tabular}{c|c|c|c|c}
 \hline\hline & & & & \\[-0.3cm]
 & 2D-Dirac & 2D-Rashba & 2D-Dresselhaus & 1D \\
 & & & & \\[-0.3cm] \hline & & & & \\[-0.3cm]
 $\hat{\lambda}(\mathbf{p})$ &
  $\lambda\, ( p_x\, \hat{\sigma}_x + p_y\, \hat{\sigma}_y)$ &
  $\lambda\, ( p_y\, \hat{\sigma}_x - p_x\, \hat{\sigma}_y)$ &
  $\lambda\, ( p_x\, \hat{\sigma}_x - p_y\, \hat{\sigma}_y)$ &
  $\lambda\, p_x\, \hat{\sigma}_x$ \\
 & & & & \\[-0.3cm] \hline & & & & \\[-0.3cm]
 $\mathcal{M}$
  & $\begin{pmatrix} 1 & 0 \\ 0 & 1 \\ 0 & 0 \end{pmatrix}$
  & $\begin{pmatrix} 0 & -1 \\ 1 & 0 \\ 0 & 0 \end{pmatrix}$
  & $\begin{pmatrix} 1 & 0 \\ 0 & -1 \\ 0 & 0 \end{pmatrix}$
  & $\begin{pmatrix} 1 & 0 \\ 0 & 0 \\ 0 & 0 \end{pmatrix}$ \\
  \hline\hline
\end{tabular}
\end{center}
\end{table}

All of the spin-orbit-coupling types considered in this work
can be expressed as
\begin{equation}
\hat{\lambda}(\mathbf{p}) = \lambda \sum_{a\in\{x,y,z\}}
\hat{\sigma}_a\,\, \sum_{\mu\in\{x,y\}} \mathcal{M}_{a \mu}\,
p_\mu \equiv \lambda\, \hat{\bm{\sigma}} \cdot \mathcal{M}\,
\mathbf{p} \quad ,
\end{equation}
with a velocity scale $\lambda$ measuring the
spin-orbit-coupling strength and $\mathcal{M}_{a\mu} \in\{0, \pm
1\}$. The $3\times 2$ 
matrix $\mathcal{M}$ connects spin and orbital degrees of
freedom. The particular choices for
$\mathcal{M}$ associated with each type of spin-orbit coupling
are also specified in Table~\ref{tab:soc_types}. In the
following, a versatile theoretical treatment of different
spin-orbit couplings is facilitated by introducing the 3-vector
of momentum-dependent spin splittings
\begin{equation}\label{eq:qMp}
 \mathbf{q} \equiv \begin{pmatrix} q_x\\ q_y\\ q_z \end{pmatrix}
 = \mathcal{M}\, \mathbf{p} \quad , 
\end{equation}
such that $\hat{\lambda}(\mathbf{p})=\lambda\, \hat{\bm{\sigma}}
\cdot \mathbf{q}$. The fact that $\mathcal{M}^T \mathcal{M} =
\mathbbm{1}_{2\times 2}$ for all of the 2D-type spin-orbit
couplings causes certain physical properties of these systems
to be identical and also underpins their qualitative difference
with the case of 1D-type spin-orbit coupling for which
$\mathcal{M}^T \mathcal{M} = \mathrm{diag}(1,0)$. As
$\mathbf{q}$ is fully in-plane (i.e., $q_z\equiv 0$) for the
spin-orbit-coupling types considered in this work, this will be
implicit in the formalism. In particular, $\mathbf{q}^2\equiv
q_x^2 + q_y^2$ in all mathematical expressions below.

The eigenstates $|\alpha_j, \mathbf{p}_j\rangle$ of the
single-particle Hamiltonian (\ref{eq:1body_hamiltonian}) are
labelled by the individual particle's momentum $\mathbf{p}_j$
and a helicity quantum number $\alpha_j = \pm 1$ that
distinguishes spin-split single-particle energy bands
$\epsilon_{\alpha_j}(\mathbf{p_j})$ with dispersions
\begin{equation} \label{eq:spenergy}
 \epsilon_\alpha(\mathbf{p}) = \frac{\mathbf{p}^2}{2m} +
 \alpha\, Z(\mathbf{p}) \quad .
\end{equation}
Here we introduced the effective spin-splitting energy
\begin{equation}
Z(\mathbf{p}) = \sqrt{h^2 +\lambda^2\, \mathbf{q}^2} 
\end{equation}
that is a function of $\mathbf{p}$ via the momentum-dependent
spin splitting $\mathbf{q}$ [see Eq.~(\ref{eq:qMp})].
Using the eigenbasis of $\hat{\sigma}_z$, the single-particle
eigenspinors can be written more explicitly as
\begin{equation}\label{eq:sPspinor}
 |\alpha, \mathbf{p}\rangle = \begin{pmatrix}
 e^{-i \phi/2}\, \sqrt{\frac{Z(\mathbf{p}) + \alpha\, h}{2Z(
 \mathbf{p})}}\\[0.2cm]
 \alpha\, e^{i\phi/2}\, \sqrt{\frac{Z(\mathbf{p}) - \alpha\,
 h}{2Z(\mathbf{p})}} \end{pmatrix} \quad ,
\end{equation}
with $\phi = \mathrm{arg}(q_x+iq_y)$.
The dispersion (\ref{eq:spenergy}) has its minimum at the
value\cite{Lin2011}
\begin{equation}\label{eq:spEmin}
 \epsilon^{\mathrm{min}} = \left\{ \begin{array}{cl}
 -\frac{1}{2} \left( m\, \lambda^2 + \frac{h^2}{m\, \lambda^2}
 \right) & \mbox{for $|h| \le m\, \lambda^2$} \\ - |h| &
 \mbox{for $|h| \ge m\, \lambda^2$} \end{array} \right. \quad .
\end{equation}

To address the two-particle problem, we switch to COM and
relative coordinates for the orbital motion,
\begin{equation}\label{eq:relative_coordinates}
 \mathbf{R} = \frac12(\mathbf{r}_1 + \mathbf{r}_2)
  \hspace{0.3cm} , \hspace{0.2cm}
 \mathbf{r} = \mathbf{r}_1 - \mathbf{r}_2
  \hspace{0.3cm} , \hspace{0.2cm}
 \mathbf{P} = \mathbf{p}_1 + \mathbf{p}_2
  \hspace{0.3cm} , \hspace{0.2cm}
 \mathbf{p} = \frac12\, (\mathbf{p}_1 - \mathbf{p}_2)
  \hspace{0.3cm} ,
\end{equation}
and introduce the total-spin operator
\begin{equation}
\hat{\bm\Sigma} = \frac{1}{2}\, \left( \hat{{\bm\sigma}}\otimes
\mathbbm{1} + \mathbbm{1} \otimes \hat{{\bm\sigma}} \right)
\end{equation}
whose eigenstates are the familiar singlet and triplet states
$|S\, M \rangle$ with $S\in \{0, 1\}$ and $M = -S, -S+1,\dots
S$ denoting eigenvalues of $\hat{\Sigma}_z$. It is possible to
separate off the COM kinetic energy and write the two-particle
Hamiltonian (\ref{eq:2body_hamiltonian}) in the form 
\begin{align}\label{eq:Htot}
 \hat{H} &= \frac{\mathbf{P}^2}{4m}+\hat{H}_\mathbf{P} +
 V(\mathbf{r}) \quad ,
\end{align}
where $\hat{H}_\mathbf{P}$ contains the relative-motion kinetic
energy and the spin-orbit coupling terms,  which still depend
parametrically on the COM momentum $\mathbf{P}$:
\begin{equation} \label{eq:H_P}
 \hat{H}_\mathbf{P} = \frac{\mathbf{p}^2}{m} + \lambda\,
 \mathbf{q} \cdot ( \hat{\bm{\sigma}}\otimes \mathbbm{1} -
 \mathbbm{1}\otimes \hat{{\bm\sigma}}) + 2\,
 \mathbf{B}_\mathbf{P}\cdot \hat{\bm\Sigma} \quad .
\end{equation}
Here we have introduced the abbreviation
\begin{align}\label{eq:BP}
\mathbf{B}_\mathbf{P} = \begin{pmatrix}
\lambda\, Q_x/2 \\
 \lambda\, Q_y/2 \\
 h
\end{pmatrix} \,\, ,
\end{align}
where $\mathbf{Q} = \mathcal{M}\, \mathbf{P}$ with the  matrix
$\mathcal{M}$ associated with the spin-orbit-coupling type as
per Table~\ref{tab:soc_types}.
The last term in Eq.~(\ref{eq:H_P}) constitutes a
Zeeman-splitting-like two-particle energy contribution, where
$\mathbf{B}_\mathbf{P}$ plays the role of an effective
three-dimensional magnetic-field vector with in-plane components
arising from the COM motion through spin-orbit coupling.
However, it is not the only term that determines the
spin-dependence of the two-particle energies and eigenstates.
As $[ \hat{\bm\Sigma}^2 , \, \hat{H}_\mathbf{P} ]\ne 0$ due to
the second term in Eq.~(\ref{eq:H_P}), two-particle eigenstates
will generally be superpositions of the eigenstates $|S\, M
\rangle$ for total spin when $\lambda\ne 0$. The form of
$\mathbf{B}_\mathbf{P}$ indicates that, when spin-orbit coupling
is finite, COM momentum affects the relative motion via a
Zeeman-like coupling to the in-plane total-spin components.
We will see below that there are certain similarities between
how finite Zeeman energy $h$ and finite COM momentum
$\mathbf{P}$ affect the two-particle binding energy, and how
they both drive the dimer state to have predominantly triplet
character when their respective Zeeman-splitting magnitudes
$|h|$ and $\lambda\, |\mathbf{Q}|$ are large. However, the
detailed bound-state structure is strongly influenced by the
interplay of $h$ and $\lambda\,\mathbf{Q}$, such that it differs
markedly in the two limits when either $\mathbf{P}$ or $h$
vanish.

We now proceed to solve the relative-motion problem embodied by
the Hamiltonian $\hat{H}_\mathbf{P} + V(\mathbf r)$.
While the coordinate transformation to relative and COM
coordinates of Eq.~\eqref{eq:relative_coordinates} has not
completely removed the dependence on COM properties, it
nevertheless reduces the dimensionality from four to two degrees
of freedom. As the total momentum $\mathbf{P}$ is a good quantum
number, the remaining $\mathbf{P}$ dependence in the
relative-motion problem is solely of parametric nature as
$\mathbf{P}$ can be considered to have a fixed value. The
$\mathbf{P}$-dependent terms in the relative-motion Hamiltonian
of Eq.~\eqref{eq:H_P} are proportional to the spin-orbit
coupling strength $\lambda$ and thus originate directly from the
spin-orbit coupling. Any of the different forms of spin-orbit
coupling considered in this work (see Table \ref{tab:soc_types})
generate such terms. We first consider the situation of
noninteracting particles and then solve the two-particle
bound-state problem.

\subsection{Case of noninteracting particles}
\label{subsec:2A}

In the absence of interactions, the two-particle eigenstates
are eigenstates of $\hat{H}_\mathbf{P}$, which can be written as
antisymmetrized products of the individual particles' helicity
and momentum eigenstates;
\begin{equation}\label{eq:HPeigen}
 |\mathbf{p}; \alpha_1, \alpha_2\rangle\rangle_\mathbf{P} =
 \mathcal{A}\, \big[|\mathbf{p})\,\, |\alpha_1, \alpha_2
 \rangle_{\mathbf{p}, \mathbf{P}}\big] \equiv \frac{1}{\sqrt{2}}
 \left( |\mathbf{p})\,\, |\alpha_1, \alpha_2
 \rangle_{\mathbf{p}, \mathbf{P}} \,\, -\,\, |-\mathbf{p})\,\,
 |\alpha_2, \alpha_1 \rangle_{\mathbf{-p}, \mathbf{P}} \right)
 \quad .
\end{equation}
Here, $|\mathbf{p})$ denotes a relative-momentum eigenstate,
$\alpha_j=\pm 1$,
and $\mathcal{A}$ is the antisymmetrization operator. Throughout
this paper, we use a notation where $|\cdot)$ denotes states for
the relative-orbital-motion degree of freedom, $|\cdot\rangle$
are spin states, and $|\cdot\rangle\rangle$ are full
two-particle states in the product space of spin and orbital
degrees of freedom. The two-particle spin state $|\alpha_1,
\alpha_2\rangle_{\mathbf{p}, \mathbf{P}}$ is a product state of
single-particle helicity states given in
Eq.~(\ref{eq:sPspinor});
\begin{equation}\label{eq:heliProd}
 |\alpha_1, \alpha_2\rangle_{\mathbf{p}, \mathbf{P}} = 
 |\alpha_1, \mathbf{P}/2 + \mathbf{p}\rangle\otimes
 |\alpha_2, \mathbf{P}/2 - \mathbf{p}\rangle \quad .
\end{equation}
The associated two-particle eigenenergies are
\begin{equation} \label{eq:noninteractingenergy}
 \varepsilon_\mathbf{P}(\alpha_1, \alpha_2, \mathbf p) =
 \frac{\mathbf{p}^2}{m} + \alpha_1\, Z_+ + \alpha_2\, Z_-
 \quad ,
\end{equation}
with the definitions
\begin{equation}
 Z_\pm = \sqrt{h^2 + \lambda^2 \left( \frac{\mathbf{Q}}{2}
 \pm \mathbf{q} \right)^2} \quad .
\end{equation}
The relative-motion energy dispersion
(\ref{eq:noninteractingenergy}) has its minimum at
\begin{equation}\label{eq:2pEmin}
 \epsilon_\mathbf{P}^{\mathrm{min}} = \left\{
 \begin{array}{cl} - m\, \lambda^2 -
 \frac{\mathbf{B}_\mathbf{P}^2}{m \lambda^2} & \mbox{for
 $\sqrt{\mathbf{B}_\mathbf{P}^2} \le m\, \lambda^2$}\\[0.1cm]
 - 2\, \sqrt{\mathbf{B}_\mathbf{P}^2} & \mbox{for
 $\sqrt{\mathbf{B}_\mathbf{P}^2} \ge m\, \lambda^2$} \end{array}
 \right. \quad .
\end{equation}
For fixed $\mathbf{P}$ and $\mathbf{p}$, the states $|\alpha_1,
\alpha_2\rangle_{\mathbf{p}, \mathbf{P}}$ form 
an orthonormal basis within
two-particle spin-ket space, thus providing a resolution of the
unity operator $\mathbbm{1}_{\mathbf{p}, \mathbf{P}}$ in this
subspace;
\begin{equation}\label{eq:unity_pP}
\sum_{\alpha_1,\alpha_2} \,\, |\alpha_1,\alpha_2
\rangle_{\mathbf{p}, \mathbf{P}}\,\, _{\mathbf{p}, \mathbf{P}}
\langle\alpha_1, \alpha_2| = \mathbbm{1}_{\mathbf{p},
\mathbf{P}} \,\, .
\end{equation}

\subsection{Bound states resulting from \textit{s}-wave
attraction}
\label{subsec:2B}

The treatment of the 2D-fermion bound-state problem
with short-range interactions in the absence of
spin-orbit coupling is well-established~\cite{Randeria1990,
Pricoupenko2011}. Recent generalizations~\cite{He2012,
Zhang2012a,Takei2012} were developed to explore ramifications of
2D-type spin-orbit coupling. Here we extend the Green's-function
formalism employed in Refs.~\cite{Zhang2012a,Takei2012} to study
the combined effects of spin-orbit coupling and Zeeman spin
splitting.

A bound state is a solution of the Schr\"odinger equation
\begin{equation}\label{eq:2pSE}
 \left[\hat{H}_{\mathbf{P}}+ V(\mathbf r)\right]|\psi_\mathrm{b}
 \rangle\rangle = E_\mathrm{b} |\psi_\mathrm{b}\rangle\rangle 
\end{equation}
with energy below the continuum of energies available to two
unbound particles. Depending on whether or not different
COM-motion states are accessible to the two-particle system
under consideration, two possible threshold energies for bound
states can be defined. In situations where dissociation can
involve transitions between different COM momenta, stability of
bound states requires their total two-particle energy
$E_\mathrm{b} + \mathbf{P}^2/(4m)$ to be below the
lowest-possible energy $2\epsilon^{\mathrm{min}}$ two unbound
particles can have, where $\epsilon^{\mathrm{min}}$ is given
by Eq.~\eqref{eq:spEmin}. In this case, one should be looking
for eigenstates $|\psi_\mathrm{b}\rangle\rangle$ of
$\hat{H}_{\mathbf{P}} + V(\mathbf{r})$ that satisfy
$E_\mathrm{b} < E_\mathrm{th}^{\mathrm{abs}}$ with
\begin{equation}\label{eq:EthAbs}
 E_\mathrm{th}^{\mathrm{abs}} = 2\, \epsilon^{\mathrm{min}}
 - \frac{\mathbf{P}^2}{4 m} \equiv \left\{ \begin{array}{cl}
 - m\, \lambda^2 - \frac{h^2}{m\, \lambda^2} -
 \frac{\mathbf{P}^2}{4 m} & \mbox{for $|h| \le m\, \lambda^2$}
 \\[0.1cm] -2\, |h| - \frac{\mathbf{P}^2}{4 m} & \mbox{for $|h|
 \ge m\, \lambda^2$} \end{array} \right. \quad .
\end{equation}
Alternatively, if the COM motion of the two-particle system is
considered to be conserved, we can focus only on the
relative-motion dynamics for a two-particle system with fixed
COM momentum $\mathbf{P}$. Then the threshold energy for bound
states is given by the minimum energy
$\epsilon_\mathbf{P}^{\mathrm{min}}$ available to the relative
motion of two unbound particles at the fixed COM momentum
$\mathbf{P}$ [see Eq.~(\ref{eq:2pEmin})], i.e., the bound states
need to satisfy $E_\mathrm{b} < E_\mathrm{th}^{\mathrm{rel}}$
with
\begin{equation}\label{eq:EthRel}
 E_\mathrm{th}^{\mathrm{rel}} =
 \epsilon_\mathbf{P}^{\mathrm{min}} \equiv \left\{
 \begin{array}{cl} - m\, \lambda^2 - \frac{h^2}{m\, \lambda^2}
 - \frac{\mathbf{Q}^2}{4 m} & \mbox{for $\sqrt{h^2 +
 \frac{\lambda^2 \mathbf{Q}^2}{4}} \le m\, \lambda^2$}\\[0.2cm]
 - 2 \sqrt{h^2 + \frac{\lambda^2 \mathbf{Q}^2}{4}} & \mbox{for
 $\sqrt{h^2 + \frac{\lambda^2 \mathbf{Q}^2}{4}} \ge m\,
 \lambda^2$} \end{array} \right. \quad .
\end{equation}
All three energies $E_\mathrm{b}$,
$E_\mathrm{th}^{\mathrm{abs}}$ and
$E_\mathrm{th}^{\mathrm{rel}}$ depend parametrically on the
COM momentum $\mathbf{P}$, and $E_\mathrm{th}^{\mathrm{abs}}
= E_\mathrm{th}^{\mathrm{rel}}$ for $\mathbf{P}=\mathbf{0}$.
For 2D-type spin-orbit coupling and $\mathbf{B}_\mathbf{P}^2
\le m^2\lambda^4$, $E_\mathrm{th}^{\mathrm{abs}}$ and
$E_\mathrm{th}^{\mathrm{rel}}$ are identical even when
$\mathbf{P}\ne \mathbf{0}$, but $E_\mathrm{th}^{\mathrm{abs}}
< E_\mathrm{th}^{\mathrm{rel}}$ when $\mathbf{P}\ne
\mathbf{0}$ for 1D-type spin-orbit coupling and/or
$\mathbf{B}_\mathbf{P}^2 \ge m^2\lambda^4$. In the present work,
we adopt $E_\mathrm{th}^{\mathrm{rel}}$ as the threshold
energy to determine the existence of two-particle bound states
and to calculate their binding energy
\begin{equation}\label{eq:bindEnDef}
\epsilon_\mathrm{b}\equiv E_\mathrm{th}^{\mathrm{rel}} -
E_\mathrm{b} \,\, .
\end{equation}
Such states are only metastable when
$E_\mathrm{th}^{\mathrm{abs}}\le E_\mathrm{b}<
E_\mathrm{th}^{\mathrm{rel}}$, but they can still be
sufficiently long-lived, and therefore accessible
experimentally, in situations when COM-changing processes are
weak. Note that both threshold energies are negative,
$E_\mathrm{th}^{\mathrm{abs}}\le E_\mathrm{th}^{\mathrm{rel}}\le
0$, and thus $E_\mathrm{b}<0$. At the same time, the binding
energy is defined to be positive; $\epsilon_\mathrm{b}>0$.

The Schr\"odinger equation (\ref{eq:2pSE}) can be formally
solved via
\begin{equation}\label{eq:SEsol}
 |\psi_\mathrm{b}\rangle\rangle = \frac{1}{E_\mathrm{b} -
 \hat{H}_{\mathbf{P}}}\, V\, |\psi_\mathrm{b}\rangle\rangle
 \quad ,
\end{equation}
as the denominator on the right-hand side is never zero
because the bound-state energy is outside the eigenvalue
spectrum of $\hat{H}_{\mathbf{P}}$. We can expand the full
bound-state wave function with respect to the relative-momentum
eigenbasis,
\begin{equation}\label{eq:psiBexp}
 |\psi_\mathrm{b}\rangle\rangle = \int\frac{d^2 p'}{(2\pi
 \hbar)^2} \,\, (\mathbf{p'}|\psi_\mathrm{b}\rangle\rangle\,\,|
 \mathbf{p'}) \quad ,
\end{equation}
keeping in mind that the expansion "coefficients" $(\mathbf{p'}|
\psi_\mathrm{b}\rangle\rangle\equiv |\psi_\mathrm{b}(\mathbf{p'}
)\rangle$ are actually still kets in spin space that
parametrically depend on the total momentum $\mathbf{P}$
and are nonorthogonal for different arguments $\mathbf{p}'$.
Inserting the expansion (\ref{eq:psiBexp}) on the r.h.s.\ of
Eq.~(\ref{eq:SEsol}) and projecting both sides onto
$(\mathbf{p}|$, we obtain
\begin{equation}\label{eq:genWF}
 |\psi_\mathrm{b}(\mathbf{p})\rangle = \frac{1}{E_\mathrm{b} -
 \hat{H}_{\mathbf{P}}}\, \int\frac{d^2p'}{(2\pi\hbar)^2}\,\,
 (\mathbf{p}|V|\mathbf{p'})\,\, |\psi_\mathrm{b}(\mathbf{p'})
 \rangle \equiv \hat{G}_\mathbf{P}(E_\mathrm{b}, \mathbf{p})\,
 \int\frac{d^2p'}{(2\pi\hbar)^2}\,\, (\mathbf{p}|V|\mathbf{p'})
 \,\, |\psi_\mathrm{b}(\mathbf{p'})\rangle \,\, ,
\end{equation}
where the Green's function $\hat{G}_\mathbf{P}(E, \mathbf{p})$
is an operator (a $4\times4$ matrix) in two-particle spin space.

A general isotropic interaction potential can be expanded in
partial waves, yielding
\begin{equation}\label{eq:intPartWav}
(\mathbf{p}|V|\mathbf{p'}) = \sum_{l=-\infty}^{\infty}
V_l(\mathbf{p},\mathbf{p'})\,\, e^{i l(\phi_\mathbf{p}-
\phi_\mathbf{p'})} \quad .
\end{equation}
Here $\phi_\mathbf{p}$ is the polar angle of the vector
$\mathbf{p}$. Furthermore, as the eigenstates $|S\, M\rangle$ of
total spin form a basis in two-particle spin space, we can
expand
\begin{equation}\label{eq:psiExpand}
|\psi_\mathrm{b}(\mathbf{p})\rangle = \sum_{S, M}\,\, |S\, M
\rangle\langle S\, M |\psi_\mathrm{b}(\mathbf{p})\rangle
\end{equation}
where, due to the antisymmetry requirement of two-fermion wave
functions, $\langle 0\, 0 |\psi_\mathrm{b}(\mathbf{p})\rangle$
must be an even function of $\mathbf{p}$, whereas the functions
$\langle 1\, M |\psi_\mathrm{b}(\mathbf{p})\rangle$ must be
odd. Inserting both (\ref{eq:intPartWav}) and
(\ref{eq:psiExpand}) into the r.h.s.\ of (\ref{eq:genWF}) yields
\begin{equation}\label{eq:WFintermed}
|\psi_\mathrm{b}(\mathbf{p})\rangle = \sum_{S, M}\,\,
\hat{G}_\mathbf{P}(E_\mathrm{b}, \mathbf{p})\, |S\, M\rangle
\,\, \sum_l\,\, \int\frac{d^2p'}{(2\pi\hbar)^2}\,\,
V_l(\mathbf{p},\mathbf{p'})\,\, e^{i l(\phi_\mathbf{p} -
\phi_\mathbf{p'})} \,\, \langle S\, M |\psi_\mathrm{b}
(\mathbf{p'})\rangle \,\, .
\end{equation}
For the case of short-range, low-energy, $s$-wave scattering,
$V_l(\mathbf{p},\mathbf{p'})\to V_0 \,\delta_{l,0}$ and the
integral on the r.h.s.\ of Eq.~(\ref{eq:WFintermed}) remains
finite (vanishes) for the singlet(triplet)-state contribution(s)
because the integrand is an even (odd) function of
$\mathbf{p'}$. Thus Eq.~(\ref{eq:WFintermed}) simplifies to
\begin{equation}\label{eq:WFsingl}
|\psi_\mathrm{b}(\mathbf{p})\rangle = \hat{G}_\mathbf{P}
 (E_\mathrm{b}, \mathbf{p})\, |0\, 0\rangle\,\,\, V_0 \int
 \frac{d^2p'}{(2\pi\hbar)^2}\,\, \langle 0\, 0|\psi_\mathrm{b}
 (\mathbf{p'})\rangle \quad ,
\end{equation}
and it follows that the bound-state wave function is obtained by
the action of the Green's function on the singlet state;
\begin{equation}\label{eq:momentum_wavefunction}
 |\psi_\mathrm{b}(\mathbf{p})\rangle = N_\mathbf{P}\,
 \hat{G}_\mathbf{P}(E_\mathrm{b}, \mathbf{p})\, |0\, 0\rangle
 \,\, .
\end{equation}
The modulus of the $\mathbf{P}$-dependent normalization factor
$N_\mathbf{P}$ is determined by the normalization condition for
$|\psi_\mathrm{b}\rangle\rangle$;
\begin{equation}\label{eq:normalize}
 \langle\langle \psi_\mathrm{b}|\psi_\mathrm{b}\rangle\rangle
 = \int\frac{d^2 p}{(2\pi\hbar)^2} \, \int\frac{d^2 p'}{(2\pi
 \hbar)^2} \,\, (\mathbf{p} | \mathbf{p'})\,\, \langle
 \psi_\mathrm{b}(\mathbf{p}) | \psi_\mathrm{b}(\mathbf{p}')
 \rangle = \int\frac{d^2p}{(2\pi\hbar)^2}\, \langle
 \psi_\mathrm{b}(\mathbf{p})|\psi_\mathrm{b}(\mathbf{p})\rangle
 = 1 \,\, ,
\end{equation}
where we used the orthogonality relation $(\mathbf{p} |
\mathbf{p'}) = 2\pi\hbar^2\, \delta(\mathbf{p}' - \mathbf{p})$
for relative-momentum eigenstates. The right-most equality from
Eq.~(\ref{eq:normalize}) demonstrates that a spin-space ket
$|\psi_\mathrm{b}(\mathbf{p})\rangle$ is not itself normalized
to unity. Inserting (\ref{eq:momentum_wavefunction}) and
recognizing also that $\hat{G}_\mathbf{P}(E_\mathrm{b},
\mathbf{p})$ is a Hermitian operator in two-particle spin space
yields
\begin{equation}
 |N_\mathbf{P}| = \left\{ \int \frac{d^2p}{(2\pi\hbar)^2}\,
 \langle 0\, 0 |\, [\hat{G}_\mathbf{P} (E_\mathrm{b},
 \mathbf{p})]^2\, |0\, 0\rangle \right\}^{-\frac{1}{2}} \,\, .
\end{equation}

The amplitudes $\langle S\, M |\psi_\mathrm{b}(\mathbf{p})
\rangle$ for the \textit{s}-wave-attraction-generated bound
state (\ref{eq:momentum_wavefunction}) can be neatly expressed
in terms of matrix elements of the Green's function,
\begin{equation}\label{eq:bsSMcomp}
\langle S\, M |\psi_\mathrm{b}(\mathbf{p})\rangle =
N_\mathbf{P} \,\, \langle S\, M |\, \hat{G}_\mathbf{P}
(E_\mathrm{b}, \mathbf{p})\, |0\, 0\rangle \quad ,
\end{equation}
for which we have obtained the general analytical expressions
(see Appendix \ref{app:greens_function} for details of the
derivation)
\begin{subequations}\label{eq:SM_project_all}
 \begin{align}\label{eq:singlet_projection}
  \langle 0\, 0|\hat{G}_\mathbf{P}(E_\mathrm{b}, \mathbf{p})|0\,
  0 \rangle &= -\frac{s}{d} \left( s^2 - 4 h^2 - \lambda^2\,
  \mathbf{Q}^2 \right)\equiv - \frac{s}{d} \left( s^2 - 4\,
  \mathbf B_\mathbf{P}^2 \right) \,\, , \\ 
 \label{eq:triplet_projection_1}
  \langle 1\, 0|\hat{G}_\mathbf{P}(E_\mathrm{b}, \mathbf{p})
  |0\, 0\rangle &= -\frac{2\lambda}{d} \left[ 2\lambda\, h\,
  \mathbf{Q}\cdot \mathbf{q} - i\, s \left( \mathbf{Q}\times
  \mathbf{q} \right)_z \right] \,\, , \\
 \label{eq:triplet_projection_2}
  \langle 1\, 1|\hat{G}_\mathbf{P}(E_\mathrm{b}, \mathbf{p})
  |0\, 0\rangle &= -\frac{\sqrt{2}\,\lambda}{d} \left[ \lambda^2
  \, \mathbf{Q}\cdot \mathbf{q}\,\, (Q_x - i\, Q_y) + (s^2 + 2 s
  \, h) (q_x - i\, q_y) \right] \,\, , \\
 \label{eq:triplet_projection_3}
  \langle 1\, -1|\hat{G}_\mathbf{P}(E_\mathrm{b}, \mathbf{p})
  |0\, 0\rangle &= -\frac{\sqrt{2}\, \lambda}{d} \left[
  \lambda^2 \, \mathbf{Q}\cdot \mathbf{q} \,\, (Q_x + i\, Q_y) +
  (s^2 - 2 s \, h) (q_x + i\, q_y) \right] \,\, .
\end{align}
\end{subequations}
Here we used the abbreviation
\begin{equation}\label{eq:dExpress}
 d = s^4 - 4 s^2 \left( \lambda^2 \mathbf{q}^2 + h^2 + \lambda^2
 \mathbf{Q}^2/4 \right) + 4 \lambda^4 \left( \mathbf{Q} \cdot
 \mathbf{q} \right)^2 \equiv 
 s^2 \left( s^2 - 4\, \mathbf{B}_\mathbf{P}^2 - 4 \lambda^2
 \mathbf{q}^2 \right) + 4 \lambda^4 \left( \mathbf{Q} \cdot
 \mathbf{q} \right)^2 \, ,
\end{equation}
and $s = \mathbf{p}^2/m - E_\mathrm{b}$. The expressions
\eqref{eq:bsSMcomp} to \eqref{eq:dExpress} for the bound-state
wave function extend similar expressions given in
Ref.~\cite{Takei2012} for the case with $h=0$ by fully
accounting for nonzero Zeeman spin splitting. Based on the
expansion (\ref{eq:psiExpand}) with amplitudes
(\ref{eq:bsSMcomp}), we define the fractional weights of
total-spin eigenstates in the bound-state wave function as
\begin{equation}\label{eq:spin_populations_integral}
 N_{SM} = \frac{\int d^2p\,\, \left| \langle S\, M|
 \hat{G}_\mathbf{P}(E_\mathrm{b}, \mathbf{p}) |0\, 0\rangle
 \right|^2}{\sum_{S, M}\, \int d^2p \, \left| \langle S\, M|
 \hat{G}_\mathbf{P}(E_\mathrm{b}, \mathbf{p})|0\, 0\rangle
 \right |^2} \quad .
\end{equation}

\subsection{Binding energy from the Bethe-Peierls boundary
condition}\label{subsec:2C}

The characteristic equation for the bound-state energy can be
found by projecting Eq.~\eqref{eq:WFsingl} onto the singlet
state and integrating over momentum, which yields
\begin{equation}\label{eq:genSec}
 \frac{1}{V_0} = \int\frac{d^2p}{(2\pi\hbar)^2}\,\, \langle 0\,
 0| \hat{G}_\mathbf{P}(E_\mathrm{b}, \mathbf{p}) |0\, 0\rangle
 \quad ,
\end{equation}
with the matrix element of the Green's function between singlets
given explicitly in Eq.~(\ref{eq:singlet_projection}). Although
principally correct, Eq.~(\ref{eq:genSec}) turns out to be
impractical for determining bound-state energies because the
integral on its r.h.s.\ is ultraviolet-divergent. \textit{Ad
hoc\/} cut-offs have sometimes been introduced to circumvent
this issue~\cite{Takei2012}. Here we address the problem using
the Bethe-Peierls boundary condition for a scattering wave
function in 2D.

We consider the equivalent of
Eq.~(\ref{eq:momentum_wavefunction}) in real space,
\begin{equation}
 |\psi_\mathrm{b}(\mathbf{r})\rangle = N_\mathbf{P} \,
 \hat{g}_\mathbf{P}(E_\mathrm{b}, \mathbf{r})\, |0\, 0\rangle
 \quad ,
\end{equation}
where $g(E_\mathrm{b}, \mathbf{r})$ denotes the real-space
Green's function\footnote{Note that $\hat{g}_\mathbf{P}
(E_\mathrm{b}, \mathbf{r})$ pertains to real space only for
the two particles' relative motion but is still in momentum
space at constant $\mathbf{P}$ for their COM motion.}
\begin{equation}
 \hat{g}_\mathbf{P}(E_\mathrm{b}, \mathbf{r}) = \int
 \frac{d^2p}{(2\pi\hbar)^2}\,\, e^\frac{i\mathbf{p}\cdot
 \mathbf{r}}{\hbar}\,\, \hat{G}_\mathbf{P}(E_\mathrm{b},
 \mathbf{p}) \quad .
\end{equation}
Employing the resolution of unity Eq.~(\ref{eq:unity_pP}) in
two-particle spin space in terms of eigenstates of
$\hat{H}_\mathbf{P}$, the real-space Green's function's
matrix element between singlets is found as
\begin{align}
& \langle 0\, 0 |\hat{g}_\mathbf{P}(E_\mathrm{b}, \mathbf{r})
  |0\, 0\rangle = \nonumber \\
& \hspace{1cm} \frac{1}{(2\pi\hbar)^2}\, \int d^2p\,\,
  \sum_{\alpha_1, \alpha_2}\, e^\frac{i\mathbf{p} \cdot
  \mathbf{r}}{\hbar}\,\, |\langle 00|\alpha_1, \alpha_2
  \rangle_{\mathbf{p}, \mathbf{P}}|^2 \left(
  \frac{1}{E_\mathrm{b} - \varepsilon_\mathbf{P}(\alpha_1,
  \alpha_2, \mathbf{p})} - \frac{1}{E_\mathrm{b} - \mathbf{p}^2
  /m}\right)\nonumber\\
& \hspace{2cm} +\,\, \frac{1}{(2\pi\hbar)^2}\, \int d^2p\,\,
  e^\frac{i \mathbf{p}\cdot\mathbf{r}}{\hbar}\,\,
  \frac{1}{E_\mathrm{b} - \mathbf{p}^2/m} \quad ,
\end{align}
where the first term on the right-hand side is regular and we
have separated off the second term, which diverges
logarithmically for $|\mathbf{r}|=0$. Noting that $E_\mathrm{b}<0$,
this second term evaluates explicitly to a modified Bessel function
\begin{align}
 \frac{1}{(2\pi\hbar)^2}\, \int d^2p\,\, e^\frac{i\mathbf{p}
 \cdot\mathbf{r}}{\hbar}\,\, \frac{1}{E_\mathrm{b} -
 \mathbf{p}^2/m} = -\frac{m}{2\pi\hbar^2}\, K_0( |\mathbf{r}| \,
 \sqrt{-m\, E_\mathrm{b}}/\hbar) \quad ,
\end{align}
for which the small-argument behavior is known: $K_0(\xi) =
- \gamma - \ln(\xi/2) + o(\xi)$. Thus we can expand the Green's
function in the short-range limit $|\mathbf{r}| \rightarrow 0$ as
\begin{align}\label{eq:short_range_wavefunction}
 \langle 0\, 0 |\hat{g}_\mathbf{P}(E_\mathrm{b}, \mathbf{r}) |0
 \, 0 \rangle = \frac{m}{2\pi\hbar^2}\, \left[ \ln\left( |\mathbf{r}|
 \sqrt{-m\, E_\mathrm{b}}/2 \hbar \right) + \gamma + F_\mathbf{P}
 (E_\mathrm{b}, h) + o(|\mathbf{r}|) \right] \quad ,
\end{align}
with finite spin-orbit coupling giving rise to the 
$\mathbf{r}$-independent contribution
\begin{align}
\label{eq:rhs_general}
 F_\mathbf{P}(E_\mathrm{b}, h) &= \frac{1}{m}\, \int
 \frac{d^2p}{2\pi}\,\, \sum_{\alpha_1, \alpha_2} \, |\langle 0\,
 0 |\alpha_1, \alpha_2 \rangle_{\mathbf{p}, \mathbf{P}}|^2\,\,
 \left( \frac{1}{E_\mathrm{b} - \varepsilon_\mathbf{P}(\alpha_1,
 \alpha_2, \mathbf{p})} - \frac{1}{E_\mathrm{b} - \mathbf{p}^2
 /m}\right) \quad , \nonumber \\
 &= \frac{1}{m}\, \int \frac{d^2p}{2\pi}\,\, \left( \langle 0\,
   0| \hat{G}_\mathbf{P}(E_\mathrm{b}, \mathbf{p}) |0\, 0
   \rangle + \frac{1}{s} \right) \equiv
   \frac{4\lambda^2}{m}\, \int \frac{d^2p}{2\pi}\,\,
   \left( \frac{\lambda^2 \left( \mathbf{Q} \cdot
   \mathbf{q} \right)^2 - s^2\, \mathbf{q}^2}{s\, d}\right)\, .
\end{align}
Here we made use of Eq.~\eqref{eq:singlet_projection} to
derive the last equality in Eq.~\eqref{eq:rhs_general}.

For the singlet component of the bound-state wave function, the
short-range behavior is described by the Bethe-Peierls boundary
conditions~\cite{Pricoupenko2011} which, for 2D systems, have a
logarithmic divergence
\begin{equation}\label{eq:BPbc}
 \langle 0\, 0|\psi_\mathrm{b}(\mathbf{r})\rangle =
 N_\mathbf{P}\, \langle 0\, 0| \hat{g}_\mathbf{P}
 (E_\mathrm{b}, \mathbf{r})|0\, 0 \rangle \propto
 \ln(|\mathbf{r}| /a_{2D}) + o(|\mathbf{r}|) \,\, .
\end{equation}
Unlike their 3D counterparts, in two spatial dimensions there is
no additional term added due to the spin-orbit
coupling~\cite{Zhang2012b}. Matching the Bethe-Peierls boundary
condition (\ref{eq:BPbc}) to the short-range limit of the
singlet projection for the bound-state wave function given in
Eq.~(\ref{eq:short_range_wavefunction}), we obtain the implicit
equation
\begin{equation}\label{eq:implicit_eq_binding}
\gamma + \ln(a_{2D} \sqrt{-m\, E_\mathrm{b}}/2\hbar) + 
F_\mathbf{P}(E_\mathrm{b}, h) = 0
\end{equation}
for the bound state energy $E_\mathrm{b}$. For convenience, we
parameterize the two-particle interaction strength in terms of
the energy scale $\epsilon_0 = \hbar^2/m a_{2D}^2$.
Introducing characteristic units in terms of the spin-orbit-coupling
strength $\lambda$ allows us to define the dimensionless quantities 
\begin{subequations}
\begin{align}
\tilde{E}_\mathrm{b} &= \frac{E_\mathrm{b}}{m\,\lambda^2} \,\, , \\
\tilde{h} &= \frac{h}{m\,\lambda^2} \,\, , \\
\tilde{\epsilon}_0 &= \frac{\epsilon_0}{m\, \lambda^2} \,\, .
\end{align}
\end{subequations}
In terms of these, Eq.~(\ref{eq:implicit_eq_binding}) becomes
\begin{equation}\label{eq:boundary_implicit}
 \gamma + \ln\left(\frac12\, \sqrt{\frac{-
 \tilde{E}_\mathrm{b}}{\tilde{\epsilon}_0}}\right) = 
 - F_\mathbf{P}(\tilde E_\mathrm{b}, \tilde h) \quad .
\end{equation}
In the absence of spin-orbit coupling, i.e., for $\lambda\to0$,
the r.h.s\ of Eq.~(\ref{eq:boundary_implicit}) vanishes and
$E_\mathrm{b}\to  - 4\, e^{-2\gamma}\epsilon_0$ is obtained,
reproducing the well-known result~\cite{Pricoupenko2011} for the
two-particle bound-state energy in two spatial dimensions.

\section{Bound-state properties for 2D-type spin-orbit
coupling}
\label{sec:3}

In this section, we consider the bound-state problem for the
case of 2D-type spin-orbit couplings of the Dirac, Rashba or
Dresselhaus forms (see Table~\ref{tab:soc_types}). The bound
state's energy and conditions for its existence are the same for
all three forms because they give rise to the same $F_\mathbf{P}
(E_\mathrm{b}, h)$ and $E_\mathrm{th}^\mathrm{rel}$. This is a
direct consequence of the relation $\mathcal{M}^T \mathcal{M}=
\mathbbm{1}_{2\times 2}$ holding for the three 2D-type
spin-orbit couplings, which ensures the universal forms
$\mathbf{q}^2\equiv\mathbf{p}^2$, $\mathbf{Q}^2\equiv
\mathbf{P}^2$ and $\mathbf{Q}\cdot\mathbf{q}\equiv\mathbf{P}
\cdot\mathbf{p}$ for momentum-dependent terms in the expressions
(\ref{eq:rhs_general}) and (\ref{eq:EthRel}). For the same
reason, the singlet component of the bound-state wave function
is also the same for all 2D-type spin-orbit couplings, but this
universality does not extend to the triplet components as these
are sensitive to the particular form of $\mathcal{M}$ [see
Eqs.~(\ref{eq:SM_project_all})].

\subsection{Case of zero center-of-mass momentum}
\label{subsec:3A}

\begin{figure}[t]
\centerline{%
\includegraphics[width=0.9\textwidth]{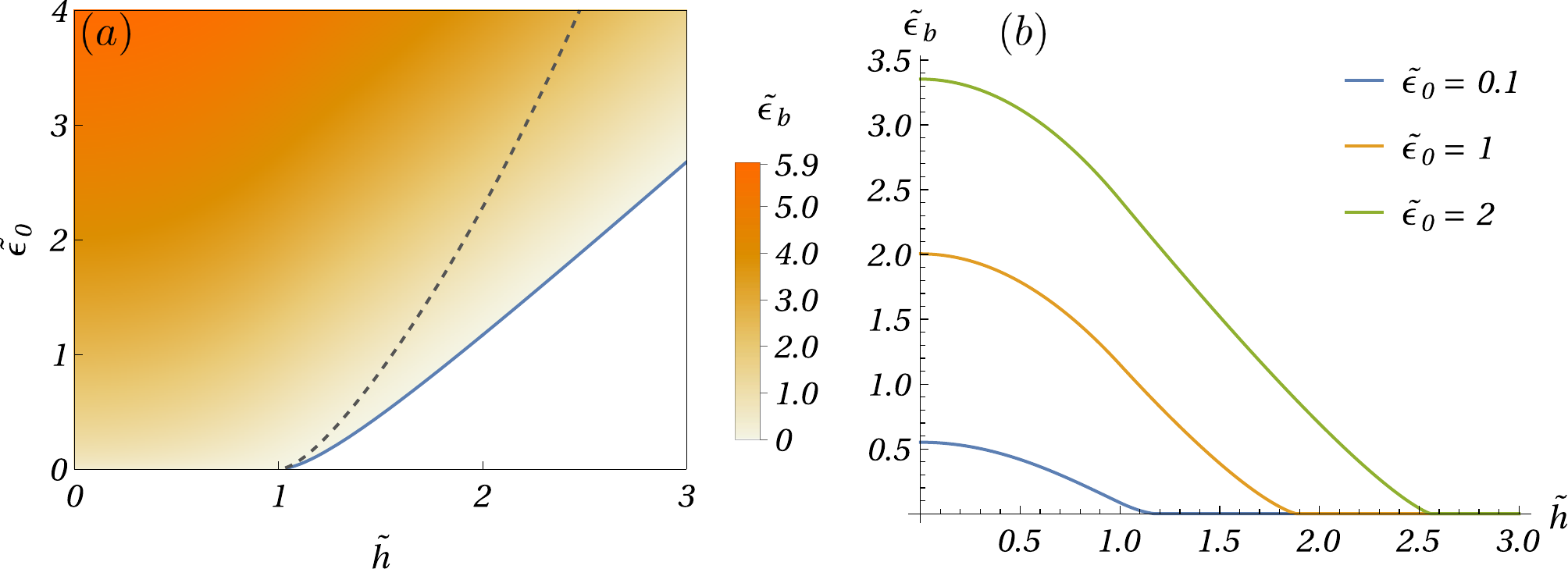}
}%
\caption{\label{fig:bound_state_phase_diagram}%
Binding energy of a 2D-fermion dimer with zero center-of-mass
momentum ($\mathbf{P} = \mathbf{0}$) formed in the presence
of 2D-type spin-orbit coupling and Zeeman splitting. Orange color
in panel~(a) indicates the parameter region where a bound state
exists. The analytical result (\ref{eq:phaseboundary}) for its
boundary is indicated by the solid blue curve, and the dashed
black curve indicates the dividing line between regions where
the two different forms of $F_\mathbf{0}(\tilde{E}_\mathrm{b},
\tilde{h})$ given in Eq.~(\ref{eq:F0forP=0}) apply. Here
$\tilde{\epsilon}_0\equiv \epsilon_0/(m\, \lambda^2)$ and
$\tilde{h}\equiv h/(m\, \lambda^2)$ are the \textit{s}-wave
interaction strength and the Zeeman energy, respectively,
measured in units of the spin-orbit-coupling energy scale $m\,
\lambda^2$. The dimensionless binding energy
$\tilde{\epsilon}_\mathrm{b} \equiv \epsilon_\mathrm{b}/(m\,
\lambda^2)$ is represented by the color scale in panel~(a) and
plotted as a function of $\tilde{h}$ for selected values
of $\tilde{\epsilon}_0$ in panel~(b).}
\end{figure}

We first assume $\mathbf{P}=\mathbf{0}$. In this case, we can
obtain an analytic expression for the quantity
$F_\mathbf{P}(\tilde{E}_\mathrm{b},\tilde{h})$ that appears on
the r.h.s.\ of Eq.~(\ref{eq:boundary_implicit});
\begin{align}\label{eq:F0forP=0}
 F_\mathbf{0}(\tilde{E}_\mathrm{b}, \tilde{h}) &=
 \frac{1}{4(\tilde{h}^2 + \tilde{E}_\mathrm{b})} \Bigg[
 - \tilde{E}_\mathrm{b} \, \ln\Bigg(
 \frac{\tilde{E}_\mathrm{b}^2}{\tilde{E}_\mathrm{b}^2 - 4
 \tilde{h}^2} \Bigg) \nonumber \\[0.2cm] & \hspace{0.5cm} -
 \left\{ \begin{array}{cl}
 \frac{2(2 \tilde{h}^2 + \tilde{E}_\mathrm{b})}{\sqrt{-1
 - \tilde{h}^2-\tilde{E}_\mathrm{b}}}\left[ \frac{\pi}{2} -
 \arctan \left( \frac{-\tilde{E}_\mathrm{b} - 2}{2 \sqrt{-1-
 \tilde{h}^2 - \tilde{E}_\mathrm{b}}} \right)\right] \Bigg] &
 \mbox{for $\tilde{E}_\mathrm{b} \le -1 - \tilde{h}^2$} \\
 \frac{2(2 \tilde{h}^2 + \tilde{E}_\mathrm{b})}{\sqrt{1 +
 \tilde{h}^2+\tilde{E}_\mathrm{b}}} \, \mbox{arcoth} \left(
 \frac{-\tilde{E}_\mathrm{b} - 2}{2\sqrt{1+\tilde{h}^2 +
 \tilde{E}_\mathrm{b}}} \right) \Bigg] & \mbox{for $-1 -
 \tilde{h}^2 < \tilde{E}_\mathrm{b} < -2 |\tilde{h}|$}
 \end{array} \right. .
\end{align}
For given dimensionless interaction strength
$\tilde{\epsilon}_0$ and Zeeman energy $\tilde{h}$,
Eq.~(\ref{eq:boundary_implicit}) constitutes an implicit
equation for the dimensionless bound-state energy
$\tilde{E}_\mathrm{b}$, which needs to be below the threshold
$\tilde{E}_\mathrm{th}\equiv E_\mathrm{th}^{\mathrm{rel}}/(m
\lambda^2)$ with $E_\mathrm{th}^\mathrm{rel}$ from
Eq.~(\ref{eq:EthRel}). We find that there is at most one such
solution of Eq.~(\ref{eq:boundary_implicit}) anywhere in parameter
space. Figure~\ref{fig:bound_state_phase_diagram} shows plots of
the dimensionless binding energy $\tilde{\epsilon}_\mathrm{b} =
\epsilon_\mathrm{b}/(m\,\lambda^2)$, with $\epsilon_\mathrm{b}$
defined in Eq.~(\ref{eq:bindEnDef}), as a function of
$\tilde{\epsilon}_0$ and $\tilde{h}$.

The nature of the bound-state solutions depends on the value of
$\tilde{h}$. For $|\tilde{h}|\le 1$ we have $\tilde{E}_\mathrm{th} =
-1-\tilde{h}^2$ according to Eq.~\eqref{eq:EthRel}. Since
$\tilde{E}_\mathrm{b}<\tilde{E}_\mathrm{th}$, the upper option
for the last term in the expression (\ref{eq:F0forP=0}) for
$F_\mathbf{0}(\tilde{E}_\mathrm{b},\tilde{h})$ applies in this case.
In particular, solving Eq.~\eqref{eq:boundary_implicit} with
$F_\mathbf{0}(\tilde{E}_\mathrm{b}, 0)$ obtained from the
$|h|\to 0$ limit of Eq.~\eqref{eq:F0forP=0} yields the result given
in Ref.~\cite{He2012} for the binding energy in the absence of
Zeeman spin splitting.

\begin{figure}[t]
\centerline{%
\includegraphics[width=0.55\textwidth]{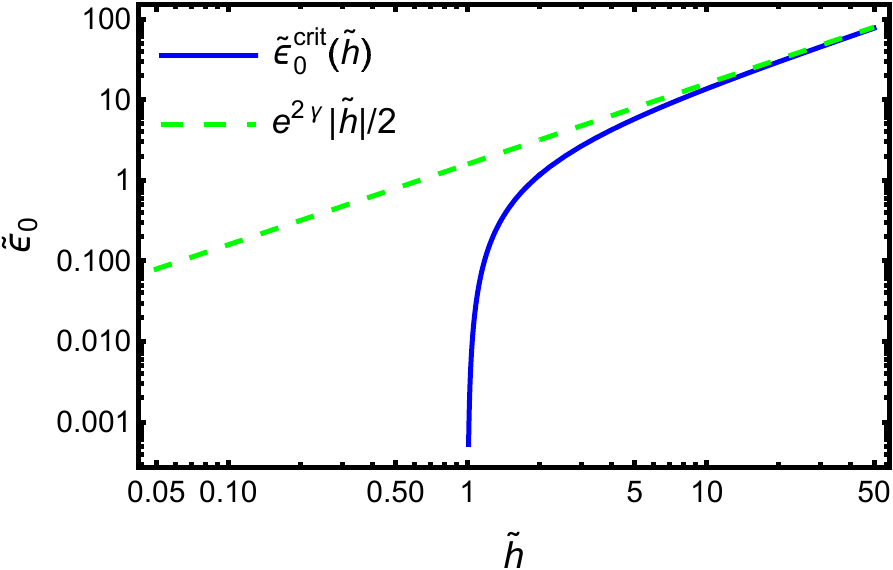}
}%
\caption{\label{fig:Clogston}%
Boundary of the region in $\tilde{\epsilon}_0$-$\tilde{h}$ space
with a two-fermion bound state. The solid blue curve plots
Eq.~\eqref{eq:phaseboundary} for the dimensionless critical
interaction strength $\tilde{\epsilon}_0^\mathrm{crit}
(\tilde{h})$ above which a bound state exists in the presence of
spin-orbit coupling. Its asymptote $e^{2\gamma}\, |\tilde{h}|/2$
for large dimensionless Zeeman coupling $\tilde{h}\equiv h/(m\,
\lambda^2)$ is indicated by the dashed green line. For pairs of
parameter values $(\epsilon_0\, , h)$ from the region above this
line, bound states are formed also in the absence of spin-orbit
coupling. In contrast, in the region between the two curves,
\textit{s}-wave-attraction-generated bound states would not
exist without spin-orbit coupling.}
\end{figure}

In contrast, for  $|\tilde{h}|>1$ we have $\tilde{E}_\mathrm{th}
= -2|\tilde{h}|$, and the region where a bound state exists has
two parts. The two parts are distinguished by whether
$\tilde{E}_\mathrm{b}\le -1- \tilde{h}^2$ or $-1-\tilde{h}^2 <
\tilde{E}_\mathrm{b} < -2|\tilde{h}|$ is satisfied and, accordingly,
which form of the last term in (\ref{eq:F0forP=0}) is applicable.
The boundary dividing these two regions can be found by letting
$\tilde{E}_\mathrm{b} \to -1 - \tilde{h}^2$ in
Eq~(\ref{eq:boundary_implicit}) with $F_\mathbf{0}
(\tilde{E}_\mathrm{b}, \tilde{h})$ from Eq.~(\ref{eq:F0forP=0})
on the r.h.s., which yields (see Appendix~\ref{app:phase_boundary}
for mathematical details)
\begin{equation}\label{eq:twoRegBound}
 \tilde{\epsilon}_0^\mathrm{div}(\tilde h)= e^{2\gamma + 2}\,\,
 \frac{1+\tilde{h}^2}{4} \left( \frac{\tilde{h}^2 -
 1}{\tilde{h}^2 + 1} \right)^{1 + \tilde{h}^2} \Theta(
 |\tilde{h}| - 1 ) \quad .
\end{equation}
Here $\Theta(\cdot)$ denotes the Heaviside step function.
In Fig.~\ref{fig:bound_state_phase_diagram}, we plot
$\tilde{\epsilon}_0^\mathrm{div}(\tilde h)$ calculated according
to Eq.~(\ref{eq:twoRegBound}) as the dashed black curve.

While a bound state always exists for small-enough Zeeman
splitting $|\tilde{h}|\le 1$, having a bound state for $|\tilde{h}|>1$
requires sufficiently strong attractive interactions. The white
area shown in Fig.~\ref{fig:bound_state_phase_diagram}(a)
indicates the parameter range for which no bound state exists.
The minimum dimensionless interaction strength
$\tilde{\epsilon}_0^\mathrm{crit}(\tilde h)$ needed to maintain
a bound state at finite Zeeman splitting is obtained by
substituting the threshold energy $\tilde{E}_\mathrm{th} =
-2 |\tilde{h}|$ applicable for $|\tilde{h}|>1$ into
Eq~(\ref{eq:boundary_implicit}), yielding (details of the derivation
are provided in Appendix \ref{app:phase_boundary})
\begin{equation}\label{eq:phaseboundary}
 \tilde{\epsilon}_0^\mathrm{crit}(\tilde h) = e^{2\gamma}\,\,
 \frac{|\tilde{h}|}{2}\, \left( 2\, \frac{|\tilde{h}|
 -1}{|\tilde{h}|} \right)^{\frac{2}{2-|\tilde{h}|}} \,
 \Theta(|\tilde{h}| - 1 ) \quad .
\end{equation}
We show $\tilde{\epsilon}_0^\mathrm{crit}(\tilde h)$ as the solid
blue line in Fig.~\ref{fig:bound_state_phase_diagram}(a). As is
apparent from Fig.~\ref{fig:bound_state_phase_diagram}(b), the
binding energy approaches zero continuously at this boundary.
In the Zeeman-splitting-dominated limit $|\tilde{h}|\gg 1$,
$\tilde{\epsilon}_0^\mathrm{crit}(\tilde h) \to e^{2\gamma}\,
\tilde{h}/2$, which is reminiscent of the Chandrasekhar-Clogston
criterion~\cite{Chandrasekhar1962,Clogston1962} for the
stability of \textit{s}-wave pairing against spin paramagnetism.
For a two-particle problem without spin-orbit coupling, the
critical interaction strength ${\epsilon}_0^\mathrm{crit}(h) =
e^{2\gamma}\, |h|/2$ emerges from equating the two-fermion binding
energy~\cite{Pricoupenko2011} $4\, e^{-2\gamma}\, \epsilon_0$
with the Zeeman-splitting energy $2|h|$.
Figure~\ref{fig:Clogston} illustrates how spin-orbit coupling
enlarges the region in parameter space where two-fermion
binding due to \textit{s}-wave attraction is possible in the
presence of Zeeman spin splitting.

\begin{figure}[t]
\centerline{%
\includegraphics[width=0.9\textwidth]{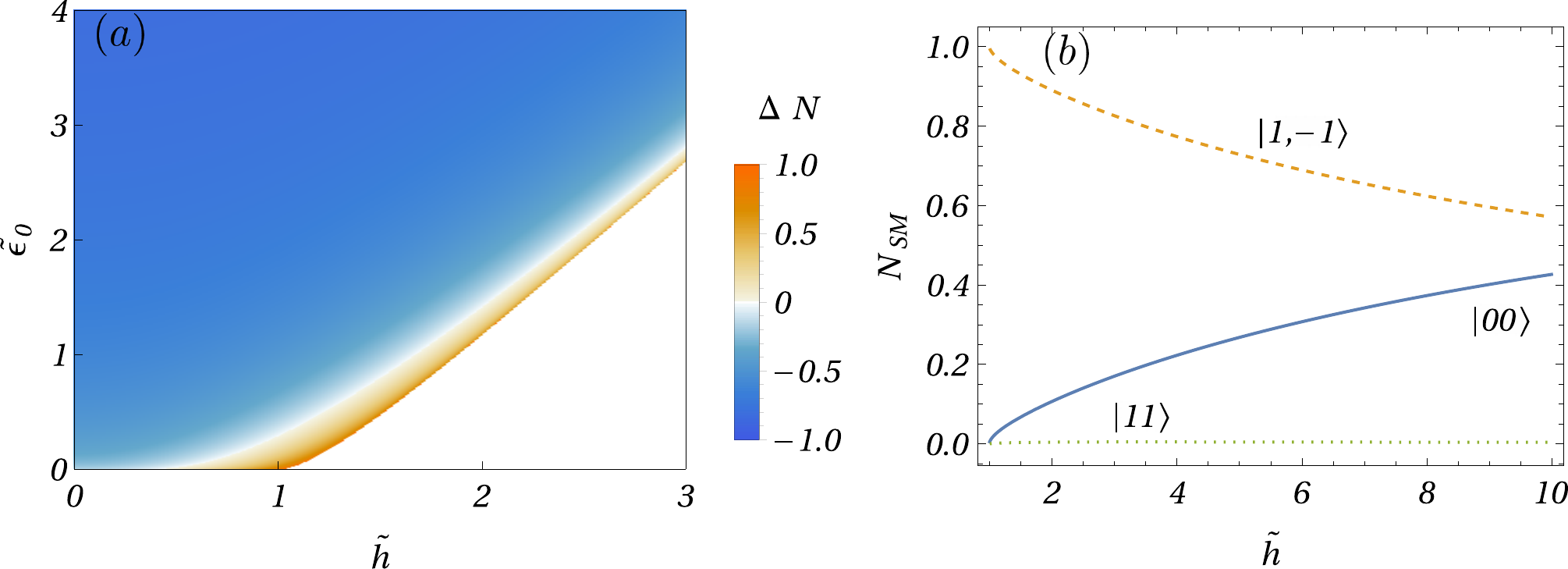}
}%
\caption{\label{fig:triplet_surplus}%
Triplet-state admixture to 2D-fermion bound states with zero
center-of-mass momentum. The color scale in panel~(a) visualizes
the difference $\Delta N = \sum_{M}  N_{1 M} - N_{0 0}$ between
the combined fractional weights of triplet states contributing
to the dimer and that of the singlet-state contribution.
Panel~(b) shows plots of $N_{S M}$ (except for $N_{1 0}=0$) for
total-spin eigenstates in bound states formed for parameter
combinations $(\tilde{h}\, ,\, \tilde{\epsilon}_0)\equiv
(\tilde{h}^\mathrm{crit} \, , \,
\tilde{\epsilon}^\mathrm{crit}_0)$ corresponding to the solid blue
line shown in Fig.~\ref{fig:bound_state_phase_diagram}(a), with
$\tilde{\epsilon}^\mathrm{crit}_0$ [$\tilde{h}^\mathrm{crit}$]
given by Eq.~(\ref{eq:phaseboundary}) [by inverting
Eq.~(\ref{eq:phaseboundary})].}
\end{figure}

It is known that the combination of \textit{s}-wave attraction
and spin-orbit coupling can result in behavior analogous to a
system subject to \textit{p}-wave interactions without
spin-orbit coupling~\cite{Zhang2008}. In our case, this would
manifest as having also triplet components of the bound-state
wave function, even though the attractive potential is
\textit{s}-wave. In the case without spin-orbit coupling,
\textit{s}-wave interactions at low energy only lead to binding
in the singlet channel, and the overlap of the wave function to
the triplet component would vanish, as the Green's function in
Eq.~(\ref{eq:momentum_wavefunction}) would be diagonal in spin
space. Turning on a Zeeman splitting $h>0$ in the absence of
spin-orbit coupling does not affect the singlet dimer until its
complete destabilization when the energy $-2\, h$ of the triplet
state $|1\, -1\rangle$ goes below the bound-state energy.
However, with spin-orbit coupling present, the \textit{s}-wave
interaction potential still projects the wave function onto the
singlet but the subsequent free propagation in the presence of
spin-orbit coupling rotates parts of the wave function back into
the triplet channel. Here we are interested to understand in
which regime a large triplet component of the bound state
develops. As the triplet state $|1\, -1\rangle$ is energetically
favored for large Zeeman splitting, it can be expected to
dominate the system. The same behavior is also seen in the
BCS mean-field theory of the many-body system where the
topological superfluid with \textit{p}-wave order parameter
emerges for large Zeeman splitting~\cite{Zhang2008,
Sato2017,Thompson2020}. As we see below, the bound-state
wave function is indeed dominated by the triplet components in
an extended region near the critical Zeeman splitting
$\tilde{h}^\mathrm{crit}(\tilde{\epsilon_0})$ obtained by
inverting the expression for $\tilde{\epsilon}_0^\mathrm{crit}
(\tilde{h})$ from Eq.~(\ref{eq:phaseboundary}).

\begin{figure}[t]
\centerline{%
\includegraphics[width=0.9\textwidth]{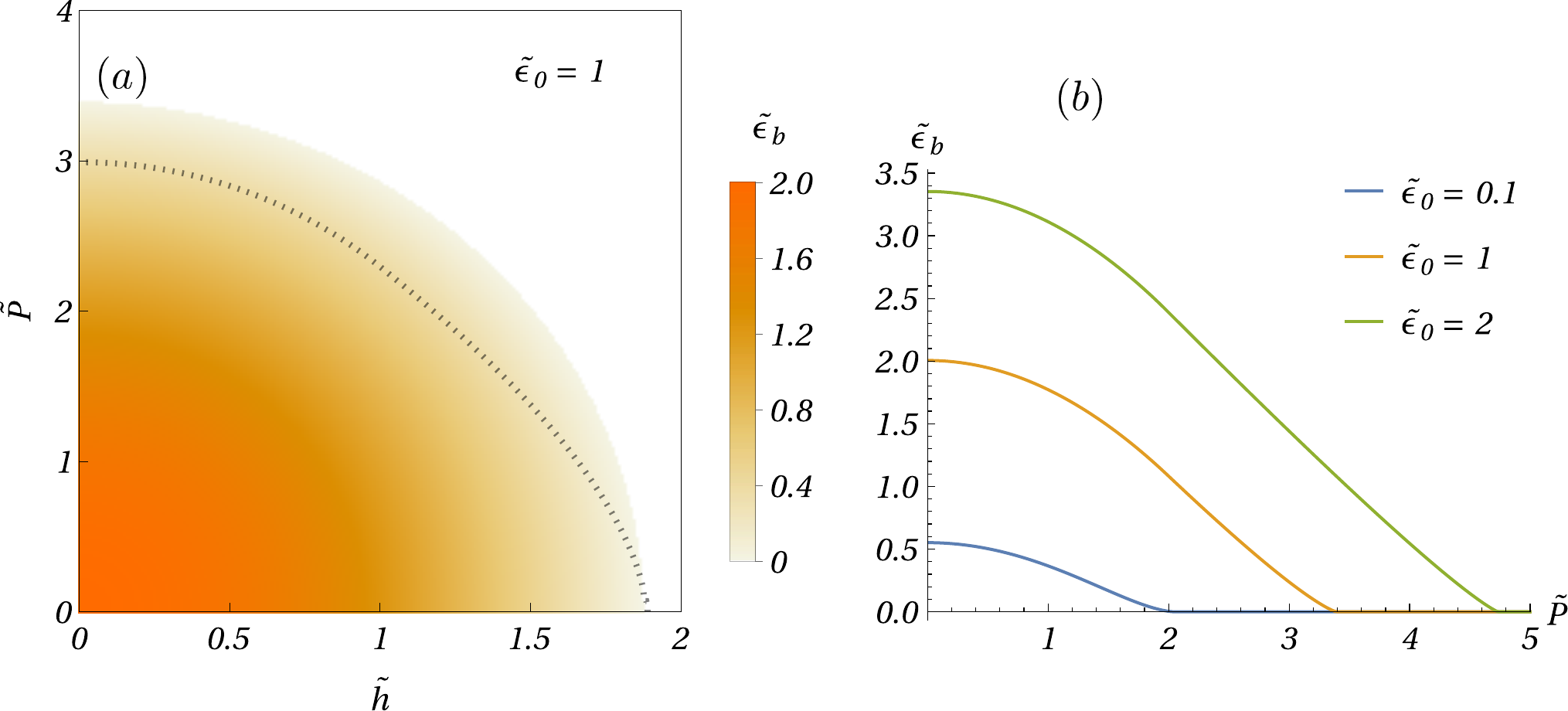}
}%
\caption{\label{fig:bound_state_finite_P}%
Effect of finite center-off-mass momentum $\mathbf{P}$ on
dimer formation. Orange color in panel~(a) indicates the range
for dimensionless parameters quantifying the COM-momentum
magnitude [$\tilde{P}\equiv|\mathbf{P}|/(m\, \lambda)$] and the
Zeeman energy [$\tilde{h}\equiv h/(m\, \lambda^2)$] within which
a 2D-fermion bound state exists. The color scale represents the
dimensionless binding energy $\tilde{\epsilon}_\mathrm{b}\equiv
\epsilon_\mathrm{b}/(m\, \lambda^2)$. The dotted black
curve delimits the region of absolute bound-state stability
where $E_\mathrm{b}< E_\mathrm{th}^\mathrm{abs} \le
E_\mathrm{th}^\mathrm{rel}$, with the threshold energies defined
in Eqs.~(\ref{eq:EthAbs}) and (\ref{eq:EthRel}). Data shown in
panel~(a) are obtained for a fixed value $\tilde{\epsilon}_0\equiv
\epsilon_0/(m\, \lambda^2)=1$ of the dimensionless interaction
strength. Panel~(b) shows plots of $\tilde{\epsilon}_\mathrm{b}$ as
a function of $\tilde{P}$ for selected values of $\tilde{\epsilon}_0$
and fixed $\tilde{h}=0$.}
\end{figure}

\begin{figure}[t]
\centerline{%
\includegraphics[width=0.9\textwidth]{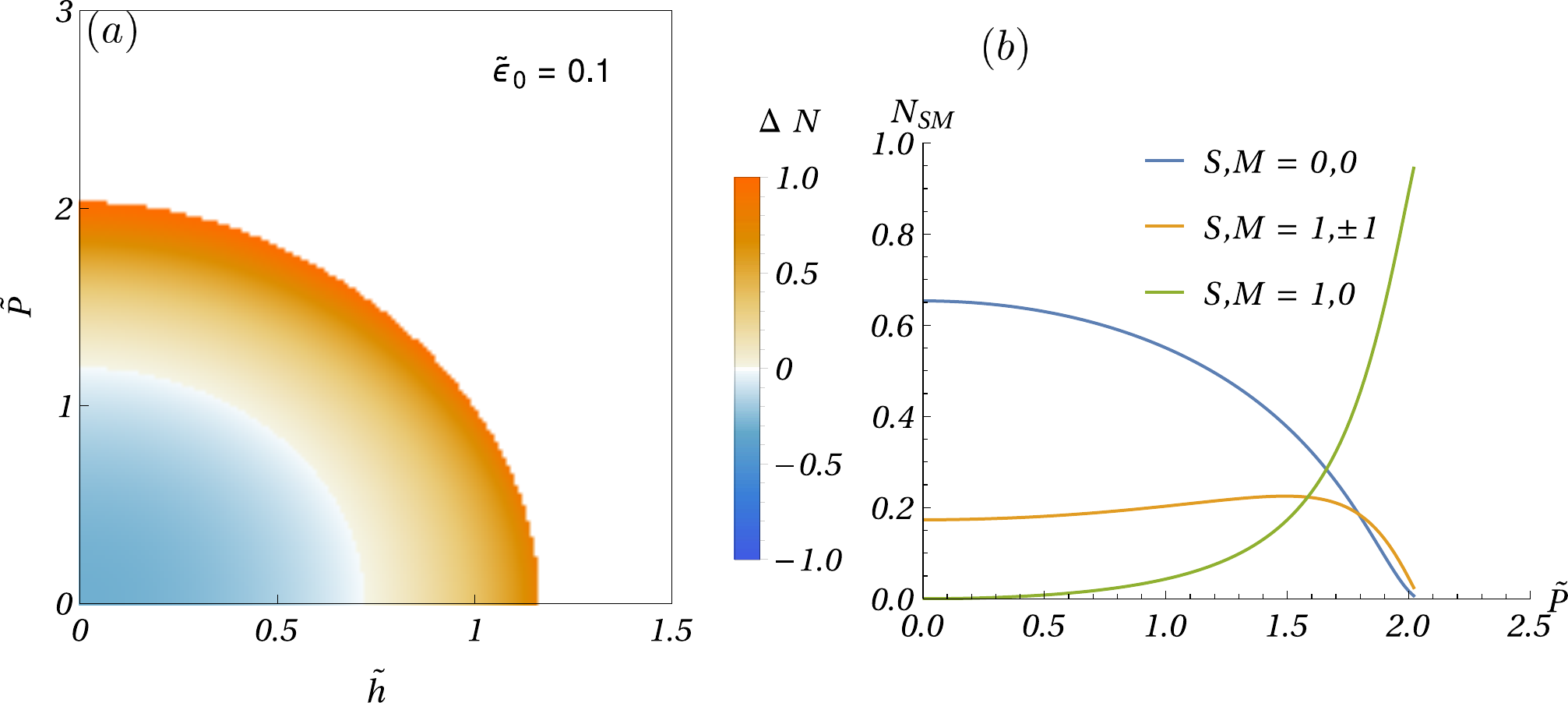}
}%
\caption{\label{fig:triplet_finite_P}%
Bound-state triplet admixture for dimers with finite
center-of-mass momentum $\mathbf{P}$. Panel~(a) shows the
quantity $\Delta N = \sum_{M}  N_{1 M} - N_{0 0}$, which
measures the balance between triplet and singlet character in
the 2D-fermion dimer, as a function of dimensionless
COM-momentum magnitude $\tilde{P}\equiv|\mathbf{P}|/(m\,
\lambda)$ and dimensionless Zeeman energy $\tilde{h} \equiv
h/(m\,\lambda^2)$. The relative weights $N_{S M}$ of individual
total-spin eigenstates contributing to the bound-state wave
function are plotted in panel~(b) as a function of $\tilde{P}$
for fixed $\tilde{h} = 0$, which is the parameter range along
the vertical axis in panel~(a). Data shown here were
calculated for fixed dimensionless interaction strength
$\tilde{\epsilon}_0\equiv \epsilon_0/(m\,\lambda^2) = 0.1$.}
\end{figure}

We obtain the fractional weights of total-spin eigenstates
contributing to the bound-state wave function via numerical
evaluation of Eq.~(\ref{eq:spin_populations_integral}), utilizing
the expressions (\ref{eq:SM_project_all}) for Green's-function
matrix elements. As the r.h.s.\ of Eq.~(\ref{eq:triplet_projection_1})
vanishes identically for $\mathbf{P}=\mathbf{0}$, the triplet state
$|1\, 0\rangle$ makes no contribution to the bound state in the
presently considered case. Figure~\ref{fig:triplet_surplus}(a) shows
the difference $\Delta N \equiv \sum_{M}  N_{1 M} - N_{0 0}$
between the combined fractional weights for triplet states and that
of the singlet state within the region of parameter space depicted in
Fig.~\ref{fig:bound_state_phase_diagram}(a). The quantity
$\Delta N$ constitutes a measure for the triplet character of
the two-particle bound state as, by construction, $-1\le
\Delta N \le 1$, where $\Delta N = 1$ indicates a pure triplet
and $\Delta N = -1$ a pure singlet state. From
Fig.~\ref{fig:triplet_surplus}(a), we see that the triplet
contribution to the bound state dwarfs the singlet part in
an extended part of parameter space adjoining the critical
boundary that delimits the region where bound states exist,
suggesting that \textit{p}-wave character of the bound-state
wave function should be prevalent there. The values $N_{S M}$
for individual triplet states are plotted in
Fig.~\ref{fig:triplet_surplus}(b) for the parameter pairs
$(\tilde{h}^\mathrm{crit}\, ,\,
\tilde{\epsilon}^\mathrm{crit}_0)$ along the boundary of the
region of existence for bound states in
Fig.~\ref{fig:bound_state_phase_diagram}(a), given explicitly by
Eq.~(\ref{eq:phaseboundary}). Asymptotically, as
$\tilde{\epsilon}_0\rightarrow 0$ and $\tilde{h}\rightarrow 1$,
the bound-state wave function becomes the state $|1\, -1
\rangle$. In contrast, the admixture of the $|1\, 1\rangle$ triplet
component to the bound-state wave function is vanishingly small,
albeit not identically zero. Thus the dimer becomes an almost
pure chiral triplet in the limit of weak interactions and
close to the critical Zeeman energy $\tilde{h}^\mathrm{crit}
\sim 1$.

\subsection{Effect of finite center-of-mass momentum}
\label{subsec:3C}

We now look for solutions of the characteristic equation
(\ref{eq:boundary_implicit}) for $\tilde{E}_\mathrm{b}$ with
finite COM momentum $\mathbf{P}$. While we were not able to
find closed-form analytic expressions for $F_\mathbf{P}
(\tilde{E}_\mathrm{b},\tilde{h})$ when $\mathbf{P}\ne
\mathbf{0}$, numerical results for the bound-state energy and
the fractional weights of the total-spin eigenstates from
Eq.~(\ref{eq:spin_populations_integral}) are readily obtained.
These turn out to not depend on the direction of $\mathbf{P}$,
as the polar angle of $\mathbf{Q}$ either does not enter
relevant mathematical expressions (e.g., the threshold
$E_\mathrm{th}^\mathrm{rel}$ is a function of $\mathbf{Q}^2$) or
can be absorbed into an integration variable (namely, the polar
angle of $\mathbf{q}$) when it explicitly appears such as
in Eq.~(\ref{eq:rhs_general}) for $F_\mathbf{P}
(\tilde{E}_\mathrm{b},\tilde{h})$.

In Fig.~\ref{fig:bound_state_finite_P}(a), the binding energy
for moderate interaction strength ($\tilde{\epsilon}_0=1$)
is plotted as a function of both the Zeeman energy $h$ and the
COM-momentum magnitude $|\mathbf{P}|$. We see that, as we
discussed earlier in this paper, COM momentum acts qualitatively
like an effective Zeeman coupling, with the region in the plane
spanned by the variables $\tilde{h} \equiv h/(m\,\lambda^2)$ and
$\tilde{P}\equiv |\mathbf{P}|/(m\, \lambda)$ where a bound state
exists exhibiting an approximately circular symmetry. The boundary
of this region is defined by the vanishing of the binding energy
(\ref{eq:bindEnDef}), i.e.,
$E_\mathrm{b} = E_\mathrm{th}^\mathrm{rel}$ with
$E_\mathrm{th}^\mathrm{rel}$ from Eq.~(\ref{eq:EthRel}). The
more stringent condition $E_\mathrm{b} <
E_\mathrm{th}^\mathrm{abs}$ with $E_\mathrm{th}^\mathrm{abs}$
given in Eq.~(\ref{eq:EthAbs}) holds within the smaller region
delimited by the dotted black curve, i.e., bound states are only
metastable within the sliver of parameter space bounded by this
curve and the boundary between orange and white regions in
Fig.~\ref{fig:bound_state_finite_P}(a).

In Fig.~\ref{fig:bound_state_finite_P}(b), we plot the
binding energy at zero Zeeman energy as a function of the
dimensionless COM-momentum magnitude $\tilde{P}$ and
observe a similar dependence as seen in
Fig.~\ref{fig:bound_state_phase_diagram}(b) as a function of the
Zeeman spin splitting. The weakening and eventual loss of the
bound state with finite COM momentum is a well known property of
dimers in the presence of spin-orbit
coupling~\cite{Vyasanakere2012,Takei2012}. It implies that,
e.g., in a not fully condensed gas with a momentum distribution
of a certain width, bound pairs at the outer edge of this
distribution are no longer bound; thus such a setup would
contain both bound pairs and unbound atoms. In a time-of-flight
measurement, the unbound atoms are expected to form a ring
around the bound pairs closer to the center of the momentum
distribution \cite{Vyasanakere2012}. 

As Fig.~\ref{fig:bound_state_finite_P}(a) shows, Zeeman
splitting and COM momentum are qualitatively similar
in their effect on the bound-state formation and its energy.
However, details of the bound-state structure are quite
different in the two cases. To illustrate this, we consider the
fractional weights of total-spin eigenstates in the bound-state
wave function, evaluating the integrals entering the expressions
(\ref{eq:spin_populations_integral}) numerically. In
Fig.~\ref{fig:triplet_finite_P}(a), we plot the difference
$\Delta N$ between the combined weights of all triplet states
and that of the singlet state as a function of $\tilde{h}$ and
$\tilde{P}$ for fixed $\tilde{\epsilon}_0=0.1$.
Triplet character is seen to be dominant all along the outer
boundary of the parameter region where dimers are formed.
However, the actual bound-state composition changes radically
as one moves between the Zeeman-energy-dominated and the
COM-momentum-dominated regimes. This becomes apparent when
comparing Fig.~\ref{fig:triplet_surplus}(b) with
Fig.~\ref{fig:triplet_finite_P}(b), where we plot the individual
fractional weights of total-spin eigenstates making up the bound
state for $\tilde{h}=0$. Here we see that, for finite COM
momenta close to the boundary of the bound-state region, the
triplet state $|1\, 0\rangle$ has the largest weight, whereas
this state does not contribute at all to the zero-COM-momentum
bound state (see Sec.~\ref{subsec:3A}). Thus the type of
dominating triplet character differs crucially depending on how
it is generated: large Zeeman splitting renders a spin-polarized
triplet state to be dominant, whereas large COM-momentum favors
the spin-unpolarized triplet state.

The behavior for large spin splitting of Zeeman or COM-motion
origin can be contrasted with the case $\tilde{P}=0$ and
$\tilde{h}=0$ that is also depicted in
Fig.~\ref{fig:triplet_finite_P}(b). In this situation, the
singlet component to the bound state is dominant and the state
$|1\, 0\rangle$ is completely absent. There is also a sizable
triplet contribution, with the oppositely spin-polarized
triplet states $|1\, \pm 1\rangle$ contributing equally to
preserve an overall spin-unpolarized bound-state wave function.
The exact values for the total-spin-eigenstate proportions in
the bound state for both COM momentum and Zeeman energy being
zero, as well as their change as the Zeeman energy becomes
finite, can be gleaned from analytical results provided in
Appendix~\ref{app:spin_populations}.

\section{Bound states formed with 1D-type spin-orbit coupling}
\label{sec:4}

\begin{figure}[t]
\centerline{%
\includegraphics[width=0.9\textwidth]{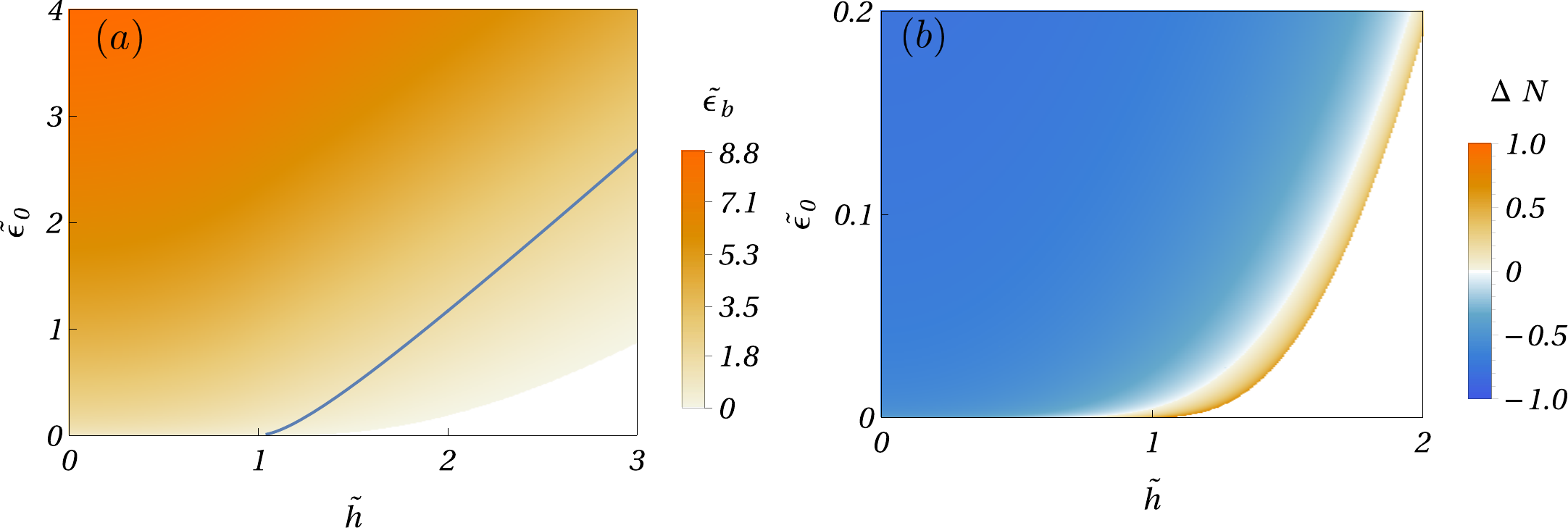}
}%
\caption{\label{fig:bound_state_phase_diagram_1d}%
Dimensionless binding energy $\tilde\epsilon_\mathrm{b}$
[panel~(a)] and bound-state triplet character quantified by
$\Delta N \equiv \sum_{M}  N_{1 M} - N_{0 0}$ [panel~(b)] of
2D-fermion dimers with zero center-of-mass momentum formed in
the presence of 1D-type spin-orbit coupling and Zeeman
splitting. These plots can be compared with corresponding
results for the 2D-type spin-orbit couplings shown in
Figs.~\ref{fig:bound_state_phase_diagram}(a) and
\ref{fig:triplet_surplus}(a), respectively. The solid
blue curve in panel~(a) is the critical-boundary line
(\ref{eq:phaseboundary}) delimiting the region where a
bound state exists for 2D-type spin-orbit coupling. Evidently,
the existence region for bound states formed with asymmetric
spin-orbit coupling extends well beyond.}
\end{figure}

The 1D-type spin-orbit coupling $\hat\lambda(\mathbf{p}) =
\lambda\, p_x\, \hat{\sigma}_x$ is not isotropic since a
particular in-plane direction is singled out. As a consequence,
the two-particle relative-motion problem only depends on
the $x$ component of the COM momentum $P_x$ through the
effective magnetic field  $\mathbf{B}_\mathbf{P}$ of
Eq.~\eqref{eq:BP}. Specifically, $P_x$ affects bound-state properties
via the dependence of $F_\mathbf{P}(\tilde{E}_\mathrm{b},\tilde{h})$,
$E_\mathrm{th}^\mathrm{rel}$ and the spinor amplitudes from
Eqs.~(\ref{eq:SM_project_all}) on $\mathbf{Q}$, whose only
nonzero component is $Q_x\equiv P_x$. The other COM component
$P_y$ is only relevant for determining the metastability
threshold $E_\mathrm{th}^\mathrm{abs}$ [see
Eq.~(\ref{eq:EthAbs})].

Specializing the general formulae from Sec.~\ref{sec:2} to the
case with 1D-type spin-orbit coupling means adopting
$\mathbf{B}_\mathbf{P} = (\lambda\, P_x/2,0,h)$, $\mathbf{Q} =
(P_x, 0, 0)$ and $\mathbf{q} = (p_x, 0, 0)$. The resulting form
of the implicit equation (\ref{eq:boundary_implicit}) for the
bound-state energy can only be solved numerically. For the case
of zero COM momentum (implying $P_x=0$), we calculate the binding
energy for the same range of Zeeman-energy values and
\textit{s}-wave-interaction strengths as in the previous
section. Results are shown in
Fig.~\ref{fig:bound_state_phase_diagram_1d}(a). We find that the
parameter region within which a bound state exists is larger
than in the case of 2D-type spin-orbit coupling. To illustrate
this, the boundary line that we derived in
Eq.~(\ref{eq:phaseboundary}) for the 2D-type case is drawn as
the solid blue curve for comparison. In addition, the
binding energy is generally higher than with 2D-type spin-orbit
coupling. 

\begin{figure}[t]
\centerline{%
\includegraphics[width=0.9\textwidth]{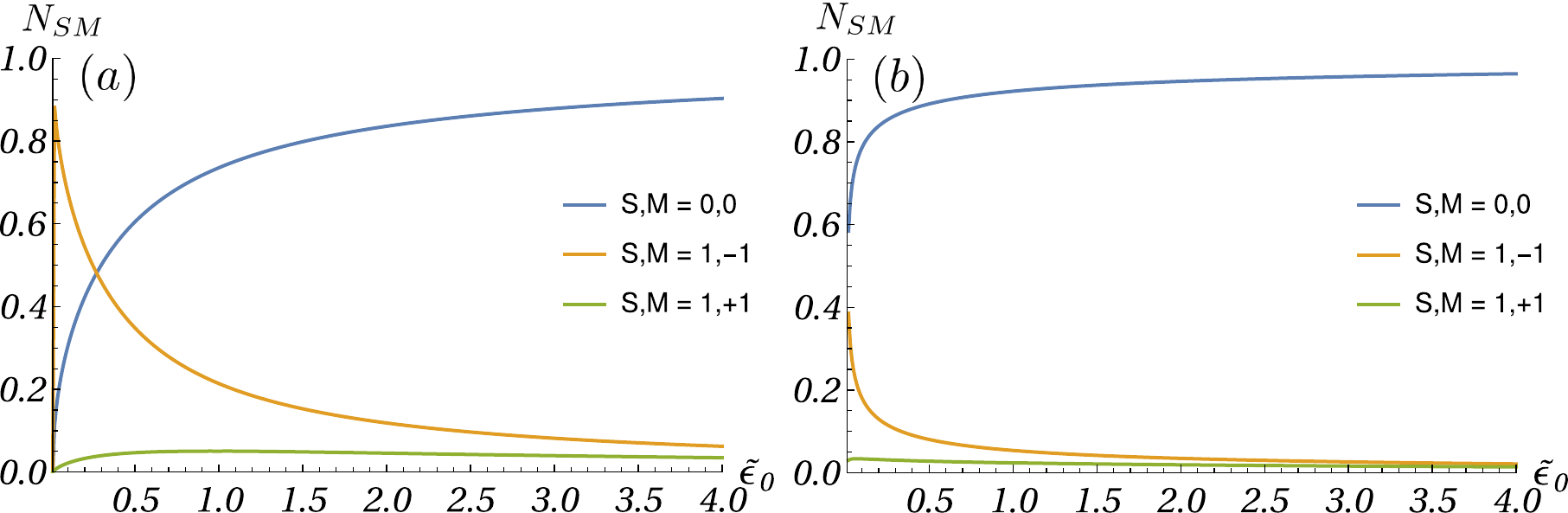}
}%
\caption{\label{fig:triplet_populations}%
Fractional weights $N_{S M}$ of total-spin eigenstates in the
2D-fermion bound state with zero COM momentum 
($\mathbf{P}=\mathbf{0}$) formed
at fixed Zeeman energy $\tilde{h} = 1$, plotted as a function of
the \textit{s}-wave interaction strength parameterized by
$\tilde{\epsilon}_0$ for 2D-type [panel (a)] and 1D-type [panel
(b)] spin-orbit couplings. We do not show $N_{1 0}$ as it
vanishes identically in both cases for zero COM momentum.}
\end{figure}

We also adapt the formalism presented in Sec.~\ref{subsec:2B}
for calculating the fractional weights $N_{S M}$ of total-spin
eigenstates in the bound state to the case of 1D-type spin-orbit
coupling. This amounts to using $\mathbf{B}_\mathbf{P} =
(\lambda\, P_x/2,0,h)$, $\mathbf{Q} = (P_x, 0, 0)$ and
$\mathbf{q} = (p_x, 0, 0)$ in Eqs.~(\ref{eq:SM_project_all}). In
Fig.~\ref{fig:bound_state_phase_diagram_1d}(b), the difference
$\Delta N$ between the total weight from triplet states contributing
to the bound state and the weight of the singlet state are
shown. The results are qualitatively similar to the case with
2D-type spin-orbit coupling [compare
Fig.~\ref{fig:triplet_surplus}(a)], but the region where the triplet
contribution to the bound state dominates has a much narrower
range in $\tilde{h}$. Figure~\ref{fig:triplet_populations} shows a
comparison in the interaction-strength dependence of relative
weights for the total-spin eigenstates present in bound states for
2D-type and 1D-type spin-orbit couplings. Again, qualitatively
similar behavior is exhibited in both cases, except that the region
of dominant triplet character occurs at much weaker interaction
strengths in the presence of 1D-type spin-orbit coupling. 
While 1D-type spin-orbit coupling is easier to realise
experimentally than the 2D types~\cite{Lin2011}, its utilization
may pose new practical challenges due to the narrower parameter
region where the triplet character dominates and the associated
smallness of binding energies (typically a fraction of the spin-orbit
energy scale $m\lambda^2$).

\section{Orbital characteristics of the bound-state wave
function}
\label{sec:5}

In previous sections, we have discussed the binding energy and
the spin properties of two-particle bound states. We now explore
features in the orbital part of the bound-state wave function.
Specifically, we focus on the amplitudes $\langle S\, M
|\psi_\mathrm{b}(\mathbf{p})\rangle$ appearing in its expansion
(\ref{eq:psiExpand}) in terms of total-spin eigenstates.

\begin{figure}[t]
\centerline{%
\includegraphics[width=0.9\textwidth]{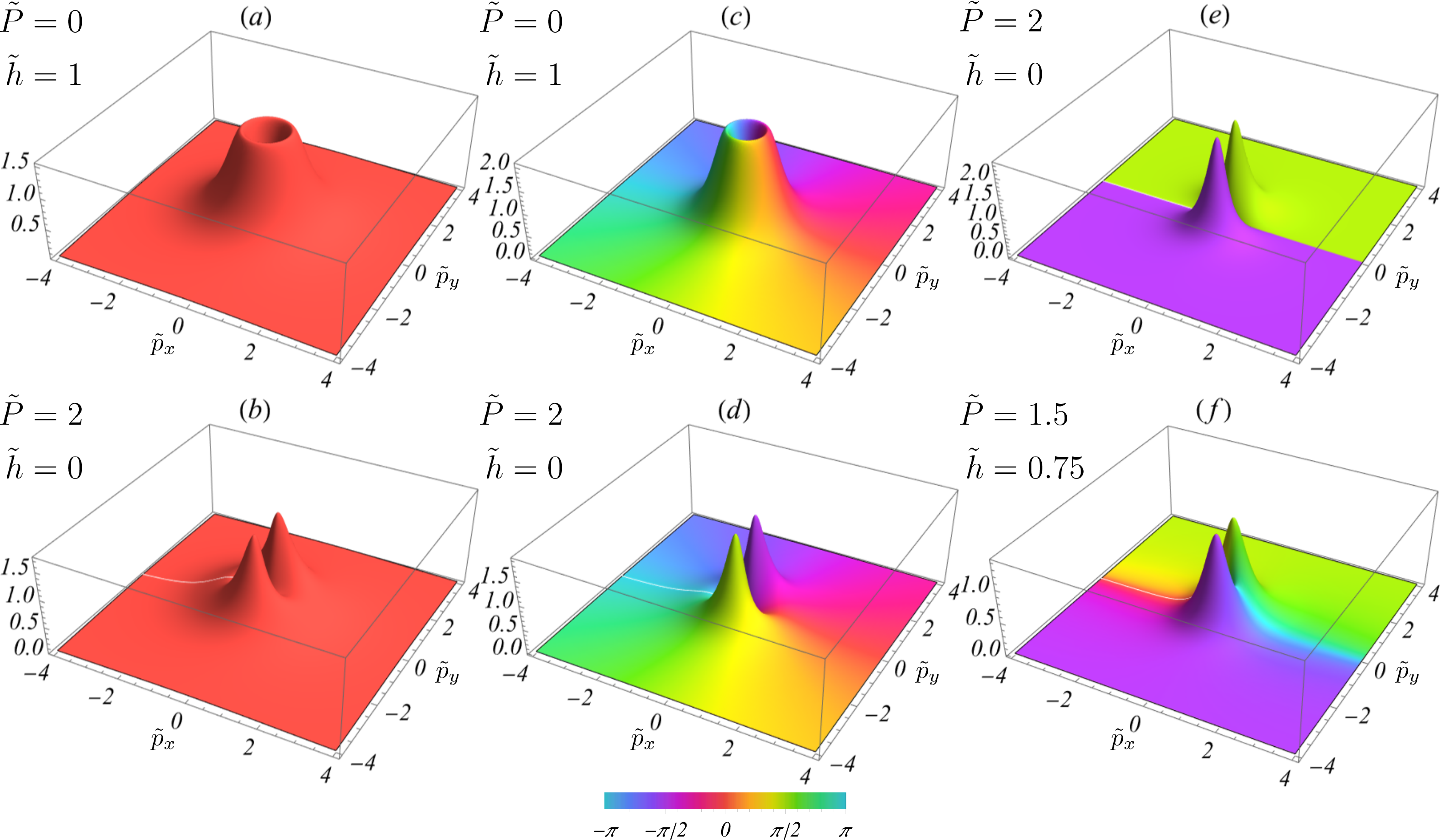}
}%
\caption{\label{fig:triplet-wf}%
Orbital wave functions of the two-fermion bound state formed in
the presence of 2D-Dirac spin-orbit coupling in
relative-momentum $\mathbf{p}\equiv (p_x, p_y)$ representation.
Surface height and color scale depict amplitude and phase,
respectively, for $\langle S\, M| \psi_\mathrm{b}(\mathbf{p})
\rangle/N_\mathbf{P} \equiv \langle S\, M | \hat{G}_\mathbf{P}
(E_\mathrm{b},\mathbf{p})\, |0\, 0\rangle$ as a function of
$\tilde{p}_j\equiv p_j/(m\,\lambda)$. Panels (a) and (b) show
the singlet component ($S=0$, $M=0$) and panels (c) and (d) the
spin-polarized triplet component ($S=1$, $M=-1$), which is
dominant for $\mathbf{P}=\mathbf{0}$, $\tilde{h}\approx 1$ and
weak \textit{s}-wave interaction strength. Panels (e) and (f)
show the spin-unpolarized triplet component ($S=1$, $M=0$),
which occurs only for finite $\mathbf{P}$. The dimensionless
interaction strength is $\tilde{\epsilon}_0=0.2$, and values
for the dimensionless Zeeman energy $\tilde{h}$ and COM momentum
$\mathbf{P}/(m\, \lambda) = (\tilde{P}, 0)$ are indicated in
each panel.}
\end{figure}

We plot representative examples obtained for 2D-Dirac spin-orbit
coupling in Fig.~\ref{fig:triplet-wf}. The singlet component,
shown in panels (a) and (b), has no imaginary part. It
exhibits radial symmetry in the relative-momentum ($\mathbf{p}$)
plane for vanishing COM momentum [panel (a)]. A local minimum
occurs at $\mathbf{p}=\mathbf{0}$ as a result of spin-orbit
coupling modifying the noninteracting dispersion
(\ref{eq:noninteractingenergy}), creating an energy minimum at
nonzero momentum. This local minimum in the orbital wave
function associated with the singlet component at vanishing
COM momentum disappears for sufficiently strong interactions
$\tilde{\epsilon}_0 \ge 1$. For finite COM momentum $\mathbf{P}$
[the case $\mathbf{P} \equiv (P\, ,\, 0)$ is shown in panel
(b)], the radial symmetry is broken as the singlet-wave-function
amplitude gets suppressed along the direction of $\mathbf{P}$.

Panels (c) and (d) in Fig.~\ref{fig:triplet-wf} show the
orbital part of the spin-polarized triplet state with $M=-1$ for
the same set of parameters used in panels (a) and (b),
respectively. This wave function is complex and, for $\mathbf{P}
=\mathbf{0}$, shows typical \textit{p}-wave behavior: a radially
symmetric amplitude and single phase winding around the node at
the origin $\mathbf{p}=\mathbf{0}$. For finite COM momentum, the
phase still behaves the same, but the radial symmetry of the
wave-function amplitude is broken in an analogous fashion as
seen for the singlet component in panel (b). Panels (e) and (f)
depict the orbital wave function for the spin-unpolarized
triplet component ($S=1$, $M=0$) that is only ever finite for
nonvanishing COM momentum. Without Zeeman splitting [panel (e)],
this wave function is purely imaginary and proportional to the
relative-momentum component perpendicular to the COM momentum
[see the cross-product term in
Eq.~(\ref{eq:triplet_projection_1})]. When the Zeeman energy is
finite as well [panel (f)], the sign change in the imaginary
part of the wave function turns into a full $2\pi$ phase
rotation, while the node along the direction parallel to
$\mathbf{P}$ softens into a finite local minimum.

\begin{figure}[t]
\centerline{
\includegraphics[width=0.9\textwidth]{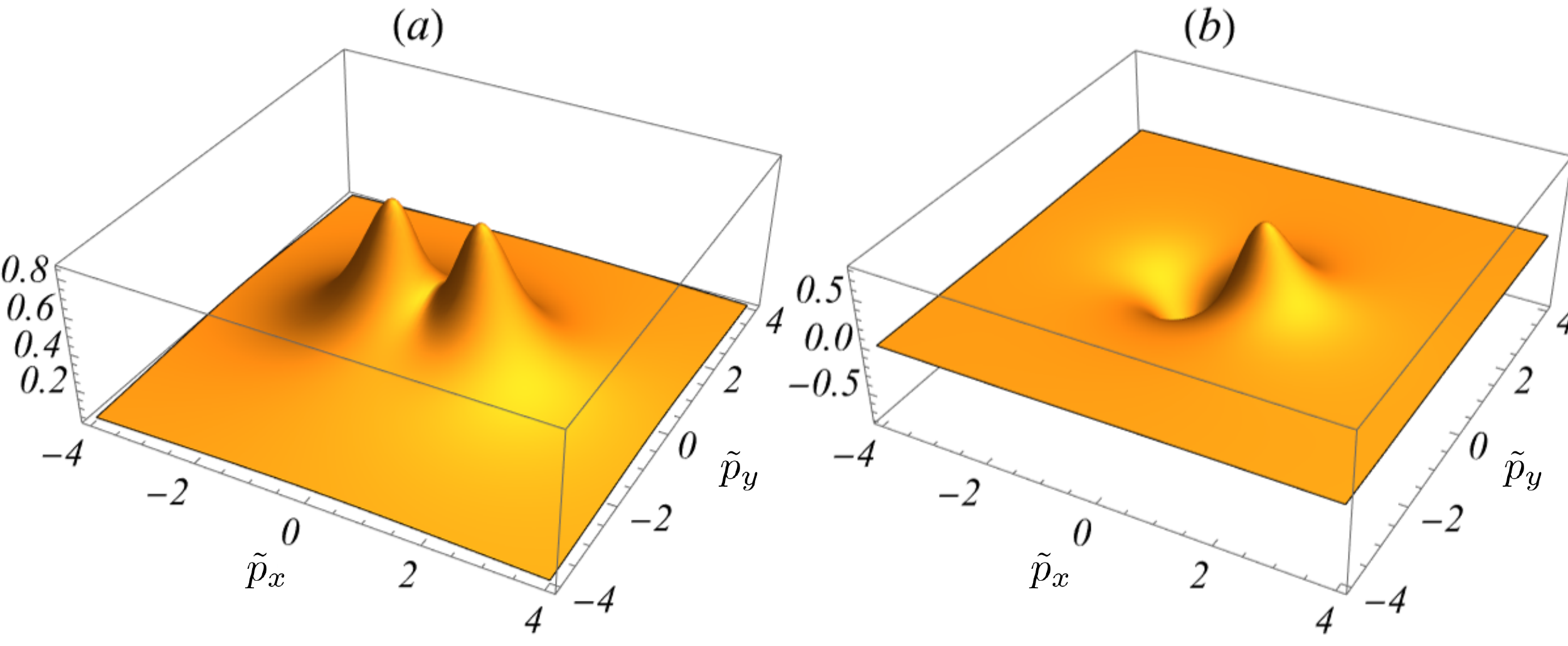}
}%
\caption{\label{fig:triplet-wf_1d}%
Orbital wave functions for two-fermion bound states formed in
the presence of 1D-type spin-orbit coupling $\hat{\lambda}
(\mathbf{p}) \equiv \lambda \, p_x\, \hat{\sigma}_x$ in the
relative-momentum $\mathbf{p}\equiv (p_x, p_y)$ representation.
Panel (a) shows the singlet $|0\,0\rangle$ component and panel
(b) the $|1\,-1\rangle$ triplet component. The surface plots
depict the real-valued functions $\langle S\, M| \psi_\mathrm{b}
(\mathbf{p}) \rangle/N_\mathbf{P} \equiv \langle S\, M |
\hat{G}_\mathbf{P}(E_\mathrm{b},\mathbf{p})\, |0\, 0\rangle$,
calculated with $\tilde{\epsilon}_0 = 0.01$ for the
dimensionless interaction strength, Zeeman splitting $\tilde h =
1$ and vanishing COM momentum $\mathbf{P}=\mathbf{0}$, in
their dependence on $\tilde{p}_j\equiv p_j/(m\,\lambda)$.}
\end{figure}

It is straightforward to adapt the results plotted in
Fig.~\ref{fig:triplet-wf} to the other 2D-type spin-orbit
couplings. According to the general formulae given in
Eqs.~(\ref{eq:SM_project_all}), the bound-state spinor
amplitudes $\langle S\, M|\psi_\mathrm{b}(\mathbf{p})\rangle$
are fundamentally a function of $\mathbf{q}$ defined in
Eq.~(\ref{eq:qMp}). As $\mathbf{q}\equiv (p_x, p_y, 0)$ for
2D-Dirac spin-orbit coupling (see the form of the matrix
$\mathcal{M}$ given for this case in Table~\ref{tab:soc_types}),
the plots from Fig.~\ref{fig:triplet-wf} in fact directly show
the momentum-space wave functions in their dependence on $q_x$
and $q_y$. Hence, the particular shape of the wave functions for
any specific 2D-type spin-orbit coupling with its associated
matrix $\mathcal{M}$ listed in Table~\ref{tab:soc_types} can be
deduced by replacing the relative-momentum components in axes
labels of plots from Fig.~\ref{fig:triplet-wf} according to the
rule $p_a\to\sum_{\mu\in\{x,y\}}\mathcal{M}_{a \mu}\, p_\mu$.
As the plots pertaining to finite $\mathbf{P}$ assumed the
particular form $\mathbf{P}=(P\, , 0)$ and therefore represent
$\mathbf{Q}=(P\, , 0, 0)$, they correspond to the case
$\mathbf{P}=\mathcal{M}^T (P,\, 0, 0)^T$ for general 2D-type
spin-orbit couplings.\footnote{This follows from multiplying both
sides of the equation $\mathbf{Q} = \mathcal{M}\, \mathbf{P}$ from
the left with $\mathcal{M}^T$ and applying the identity
$\mathcal{M}^T \mathcal{M} = \mathbbm{1}_{2\times 2}$ that holds
for all 2D-type spin-orbit couplings.}

Results shown in Fig.~\ref{fig:triplet-wf_1d} for 1D-type
spin-orbit coupling look like anisotropic versions of the
behavior seen for 2D-Dirac spin-orbit coupling. As the
amplitudes $\langle S\, M| \psi_\mathrm{b}(\mathbf{p})\rangle$
are all real-valued in this case, there is no phase winding but
simply a sign inversion in the triplet component. Instead of
a radially symmetric minimum, the singlet part exhibits a
saddle point at $\mathbf{p}=\mathbf{0}$.

\section{Experimental detection}
\label{sec:6}

Bound states could be probed with radio-frequency spectroscopy
\cite{Vale2021a}, or with high spectral resolution using
magneto-association spectroscopy \cite{Fuchs2008}.
Spin-selective imaging can provide information on the spin
content of ultra-cold atomic gases \cite{Bergschneider2018}. In
order to determine the relative weight of different total-spin
contributions to bound states, we propose to turn off the
spin-orbit-coupling fields, which projects the cold-atom
population to spin eigenstates, before using spin-selective
imaging of single-particle populations. The \textit{p}-wave
character of the bound-state wave function can be detected by
time-of-flight imaging of the single-particle momentum
distribution. The characteristic signature of the
\textit{p}-wave character of the bound state is a vortex-like
momentum distribution with a hole in the center, see
Fig.~\ref{fig:triplet-wf}. In the regime where triplet character
dominates in the bound state, a density maximum is expected at
the momentum scale $m\, \lambda\,
\sqrt{\tilde{\epsilon}_\mathrm{b} + \tilde{h}^2-1}$. 
As the square root is typically of order unity, this
yields the characteristic momentum scale of spin-orbit coupling,
which is well accessible in current experiments. The parameter
regime with a large triplet component in the bound state could
be probed by a fluorescence-imaging approach with single-atom
spin and momentum resolution, as recently
demonstrated~\cite{Holten2021}.
Due to the single-particle resolution achieved in this
experiment, it is possible to obtain relative momentum
distributions at fixed COM momentum by post selection.
The ratio of pairs with same and opposite spin would give a
clear indication on the number of singlet or triplet pairs in
the system. 

\section{Conclusions and outlook}
\label{sec:7}

In this paper, we have investigated the properties of bound
states of two fermions in a 2D gas with Zeeman spin splitting
and spin-orbit coupling. While spin-orbit coupling enhances
binding, both the Zeeman splitting and a finite COM momentum of
the dimer counteract the formation of bound states. We show that
the COM momentum acts like an additional in-plane component of
the Zeeman coupling. The bound state ceases to exist when either
or both the Zeeman energy and the COM momentum exceed a
threshold. For 1D-type spin-orbit coupling, the binding is
stronger and the Zeeman energy for which a bound state can exist
is larger than for 2D-type spin-orbit coupling.

Further, we have calculated the fractional weights of individual
total-spin components in the bound state. In the systems we
consider in this paper, there is a competition between the
\textit{s}-wave interactions, which project the two-body wave
function onto the singlet state, and the spin-orbit coupling,
which rotates the total-spin state into the triplet sector. By
this mechanism, the triplet character of the wave function can
become dominant. This happens when the Zeeman energy $|h|$ is
near the critical value for the existence of a bound state. In
this regime, the wave function is mostly in the spin-polarized
triplet state that minimizes the total energy and has a
\textit{p}-wave-like shape with a node at zero relative
momentum. We find that, for 1D-type spin-orbit coupling, this
regime where triplet states dominate occurs in a much
narrower range of Zeeman energies for fixed interaction strength
(and \textit{vice versa}) as compared to systems with 2D-type
spin-orbit coupling. Nevertheless, large triplet-state fractions
are still reached also for the bound states formed in the
presence of 1D-type spin-orbit coupling.

With finite COM momenta, we find that the bound state also
reaches dominant triplet character but now in the unpolarized
$S=1$, $M=0$ triplet state. This triplet component is only
present for nonzero COM momentum. These findings show that, in
a many-body system such as a thermal Fermi gas, the distribution
of COM momenta will lead to a gas with bound pairs in the
singlet state at the center of the momentum distribution,
triplet pairs further out, and unbound fermions at even higher
momenta. We also discuss how such bound states could be detected
experimentally; in particular, the detection of opposite-spin
and same-spin correlations can reveal whether a 2D Fermi gas
with spin-orbit coupling contains singlet or triplet bound
pairs.

\section*{Acknowledgements}
We thank Chris Vale and Paul Dyke for helpful discussions.


\paragraph{Funding information}
This work was partially supported by the
\href{https://search.crossref.org/funding?q=501100009193&from_ui=yes}{\sf Marsden Fund}
of New Zea\-land (contract nos.\ VUW1713 and MAU2007) from
government funding managed by the Royal Society Te Ap\=arangi.

\begin{appendix}

\label{app:st_amplitudes}
\section{Singlet and triplet projections of helicity-basis
product states}

This section provides useful identities involving the
two-particle states (\ref{eq:heliProd}) that are direct products
of single-particle energy eigenstates labelled by the individual
particles' momentum $\mathbf{p}_j$ and helicity $\alpha_j$.
Indicating the spin-up (spin-down) eigenstate of
$\hat{\sigma}_z$ by $|\!\uparrow\rangle$ ($|\!\downarrow
\rangle$), we get for the singlet projection of such states
\begin{align}
& \langle 0\, 0|\alpha_1, \alpha_2\rangle_{\mathbf{p},
\mathbf{P}} = \frac{1}{\sqrt{2}}\, \left( \langle\uparrow\!|
\alpha_1, \mathbf{p}_1\rangle\,\langle\downarrow\!|\alpha_2,
\mathbf{p}_2 \rangle - \langle\downarrow\!|\alpha_1,
\mathbf{p}_1 \rangle \, \langle\uparrow\!|\alpha_2, \mathbf{p}_2
\rangle \right) \nonumber\\
&= \frac{1}{\sqrt{2}} \left( \alpha_2\, e^{-\frac{i}{2} (\phi_1
- \phi_2)}\,\, \sqrt{\frac{Z_+ +\alpha_1 h}{2Z_+}}\,\,
\sqrt{\frac{Z_- -\alpha_2 h}{2Z_-}} - \alpha_1 e^{\frac{i}{2}
(\phi_1-\phi_2)}\,\, \sqrt{\frac{Z_+ -\alpha_1 h}{2Z_+}}\,\,
\sqrt{\frac{Z_- + \alpha_2 h}{2Z_-}}\right) \,\, .
\end{align}
Analogously, for the overlap with the $S=1$, $M=0$ triplet
state, we find
\begin{align}
& \langle 1\, 0|\alpha_1, \alpha_2\rangle_{\mathbf{p},
\mathbf{P}} = \frac{1}{\sqrt{2}}\left(\langle\uparrow\!|\alpha_1
, \mathbf{p}_1 \rangle\, \langle\downarrow\!|\alpha_2,
\mathbf{p}_2 \rangle + \langle\downarrow\!|\alpha_1,\mathbf{p}_1
\rangle\,\langle\uparrow\!| \alpha_2,\mathbf{p}_2\rangle \right)
\nonumber\\
&=\frac{1}{\sqrt{2}}\left(\alpha_2 e^{-\frac{i}{2} (\phi_1 -
\phi_2)}\,\, \sqrt{\frac{Z_+ +\alpha_1 h}{2Z_+}}\,\,
\sqrt{\frac{Z_- -\alpha_2 h}{2Z_-}}+\alpha_1 e^{\frac{i}{2}
(\phi_1-\phi_2)}\,\,\sqrt{\frac{Z_+ -\alpha_1 h}{2Z_+}}\,\,
\sqrt{\frac{Z_- +\alpha_2 h}{2Z_-}} \right).
\end{align}
For the projections onto the spin-polarized triplet states,
straightforward calculation yields
\begin{equation}
\langle1\, 1|\alpha_1, \alpha_2\rangle_{\mathbf{p}, \mathbf{P}}
= \langle \uparrow\!|\alpha_1,\mathbf{p}_1\rangle\,\langle
\uparrow\!|\alpha_2,\mathbf{p}_2\rangle = e^{-\frac{i}{2}
(\phi_1+\phi_2)} \,\, \sqrt{\frac{Z_+ +\alpha_1 h}{2Z_+}} \,\,
\sqrt{\frac{Z_- + \alpha_2 h}{2Z_-}}
\end{equation}
and
\begin{equation}
\langle 1\, -1|\alpha_1, \alpha_2\rangle_{\mathbf{p},
\mathbf{P}} = \langle\downarrow\!|\alpha_1,\mathbf{p}_1\rangle\,
\langle\downarrow\!|\alpha_2,\mathbf{p}_2\rangle=e^{\frac{i}{2}
(\phi_1+\phi_2)}\,\, \sqrt{\frac{Z_+ -\alpha_1 h}{2Z_+}}\,\,
\sqrt{\frac{Z_- -\alpha_2 h}{2Z_-}} \,\, .
\end{equation}
The phases appearing in these identities are $\phi_j =
\mathrm{arg}(p_{j,x} + i\, p_{j,y})$.

Relevant for calculations leading to results presented in this
paper are the absolute square of the singlet projection and the
latter's products with the triplet projections. To obtain more
compact expressions, 3D vectors $\mathbf{Z}_\pm = (\lambda
[Q_x/2\pm q_x], \lambda [Q_y/2\pm q_y], h)$ are introduced, in
terms of which we find
\begin{align}
&|\langle 0\, 0|\alpha_1, \alpha_2\rangle_{\mathbf{p},
\mathbf{P}}|^2 = \frac14\left( 1 - \alpha_1\alpha_2\,
\frac{\mathbf{Z}_+\cdot\mathbf{Z}_-}{Z_+ Z_-} \right) \,\, ,\\
&\langle 1\, 0|\alpha_1,\alpha_2\rangle_{\mathbf{p},\mathbf{P}}
\; {}_{\mathbf{p},\mathbf{P}}\langle\alpha_1,\alpha_2|0\, 0
\rangle = \frac14\left( \frac{h\, (\alpha_1\, Z_- - \alpha_2\,
Z_+)}{Z_+ Z_-} + i\, \alpha_1\alpha_2\, \frac{(\mathbf{Z}_+
\times \mathbf{Z}_-)_z}{Z_+ Z_-} \right) \,\, , \\
&\langle 1\, 1|\alpha_1,\alpha_2\rangle_{\mathbf{p},\mathbf{P}}
\; {}_{\mathbf{p},\mathbf{P}}\langle\alpha_1,\alpha_2|0\, 0
\rangle = \frac{1}{4\sqrt 2}\left(\alpha_2\, e^{-i\phi_2}\,
\frac{\lambda|\mathbf{p}_2|\, \sqrt{(Z_+ +\alpha_1\, h)^2}}{Z_+
Z_-}\right. \nonumber \\ & \hspace{8cm} \left. -\, \alpha_1 \,
e^{-i\phi_1}\,\,\frac{\lambda|\mathbf{p}_1|\,\sqrt{(Z_- +
\alpha_2\, h)^2}}{Z_+ Z_-} \right) \,\, , \\
&\langle 1\, -1|\alpha_1\,\alpha_2\rangle_{\mathbf{p},
\mathbf{P}}\; {}_{\mathbf{p},\mathbf{P}}\langle\alpha_1,\alpha_2
|0\, 0\rangle = \frac{1}{4\sqrt 2}\left( \alpha_2\, e^{i\phi_1}
\,\frac{\lambda|\mathbf{p}_1|\, \sqrt{(Z_- - \alpha_1\,
h)^2}}{Z_+ Z_-}\right. \nonumber \\ & \hspace{8cm} \left. -\,
\alpha_1\, e^{i\phi_2}\, \frac{\lambda|\mathbf{p}_2|\,
\sqrt{(Z_+ - \alpha_2\, h)^2}}{Z_+ Z_-}\right) \,\, .
\end{align}

\section{Momentum representation of the Green's function}
\label{app:greens_function}

Here, we show how to obtain somewhat compact expressions for the
momentum representation of the Green's function. As we consider
only \textit{s}-wave interactions that couple exclusively to the
singlet channel, relevant formulae always contain the Green's
function acting on the singlet total-spin eigenstate to the
right. Thus we need to calculate the four matrix elements
\begin{equation}\label{eq:greensfunction_momentum}
\langle S\, M|\hat{G}_\mathbf{P}(E,\mathbf{p})|0\, 0\rangle =
\sum_{\alpha_1, \alpha_2}\,\, \frac{\langle S\, M| \alpha_1,
\alpha_2\rangle_{\mathbf{p}, \mathbf{P}}\; _{\mathbf{p},
\mathbf{P}}\langle\alpha_1, \alpha_2|0\, 0 \rangle}{E -
\varepsilon_\mathbf{P}(\alpha_1, \alpha_2, \mathbf{p})} \quad ,
\end{equation}
where we employed the Lehmann representation in terms of the
eigenstates (\ref{eq:HPeigen}) of $\hat{H}_\mathbf{P}$.

Numerators appearing in (\ref{eq:greensfunction_momentum}) have
been obtained in the previous section. As was first shown in
Ref.~\cite{Takei2012}, introducing the variable $s=\mathbf{p}^2/m-E$
allows us to perform the sum over the four combinations of
$\{\alpha_1 = \pm, \alpha_2=\pm\}$ and obtain a more compact
form of the Green's function. For illustration, we show this in
detail for the singlet component:
\begin{align}
\langle 0\, 0|\hat{G}_\mathbf{P}(E,\mathbf p)|0\, 0\rangle &=
\sum_{\alpha_1\, \alpha_2}\,\, \frac{|\langle 0\, 0| \alpha_1,
\alpha_2\rangle_{\mathbf{p},\mathbf{P}}|^2}{- (\alpha_1 Z_+ +
\alpha_2 Z_-) - s} \\
&= \frac14 \left[ \left( 1 - \frac{\mathbf{Z}_+\cdot
\mathbf{Z}_-}{Z_+ Z_-}\right)\left(\frac{1}{Z_+ + Z_- - s} +
\frac{1}{-Z_+ - Z_- -s}\right) \right. \nonumber \\ &
\hspace{2cm} \left. +\, \left( 1 + \frac{\mathbf{Z}_+
\cdot\mathbf{Z}_-}{Z_+ Z_-}\right) \left(\frac{1}{Z_+ - Z_- -s}
+ \frac{1}{-Z_+ + Z_- -s}\right)\right]\nonumber \\
&=-\frac{s}{2}\left[\frac{1}{s^2 - (Z_+ + Z_-)^2} + \frac{1}{s^2
- (Z_+ - Z_-)^2} \right. \nonumber \\
& \hspace{2cm}\left. +\,\frac{\mathbf Z_+\cdot\mathbf{Z}_-}{Z_+
Z_-}\left(\frac{1}{s^2 - (Z_+-Z_-)^2} - \frac{1}{s^2 - (Z_+ +
Z_-)^2}\right)\right]\nonumber\\
&=-\frac{s}{d}\left[s^2 - (Z_+^2+Z_-^2-2\,\mathbf{Z}_+\cdot
\mathbf{Z}_-)\right]\nonumber\\
&=-\frac{s}{d}\left[s^2 - 4 h^2 - \lambda^2 ( q_1^2 + q_2^2 +
2\mathbf{q}_1\cdot\mathbf{q}_2)\right]\nonumber\\
&=-\frac{s}{d}\left[s^2 - 4 h^2 - \lambda^2(\mathbf{q}_1 +
\mathbf{q}_2)^2\right]=-\frac{s}{d}\left[s^2-4 h^2-\lambda^2
\mathbf{Q}^2\right]\,\, ,
\end{align}
thus obtaining (\ref{eq:singlet_projection}) with
(\ref{eq:dExpress}) giving the explicit expression for the
denominator $d$. Along the same lines, the expressions
(\ref{eq:triplet_projection_1}), (\ref{eq:triplet_projection_2})
and (\ref{eq:triplet_projection_3}) for Green's-function matrix
elements involving the triplet states are derived. Our results
agree with the expressions given in Ref.~\cite{Takei2012} for the
vanishing-Zeeman-splitting limit $|h|\rightarrow 0$.

\section{Boundaries of parameter regions for $\mathbf{P}=
\mathbf{0}$ bound states}
\label{app:phase_boundary}

For 2D-type spin-orbit coupling and zero COM momentum, the
boundary between the parameter regions with and without a bound
state can be obtained analytically. In the following, we assume
$|\tilde{h}| > 1$ since there is always a bound state when
$|\tilde{h}|\le 1$. To derive
$\tilde{\epsilon}_0^\mathrm{crit}$, we substitute the threshold
energy $\tilde{E}_\mathrm{th} = -2 |\tilde{h}|$ applicable for
$|\tilde{h}| > 1$ in place of $\tilde{E}_\mathrm{b}$ into
Eq.~(\ref{eq:boundary_implicit}), using also the analytical
expression (\ref{eq:F0forP=0}) for $F_\mathbf{0}
(\tilde{E}_\mathrm{b}, \tilde{h})$. First we consider
\begin{align}\label{eq:FP0intermed}
F_\mathbf{0}(\tilde{E}_\mathrm{b},\tilde{h})
\big|_{\tilde{E}_\mathrm{b} = -2|\tilde{h}|}\, &=
\lim_{\tilde{E}_\mathrm{b}\to -2|\tilde{h}|} \left\{
\frac{-\tilde{E}_\mathrm{b}\,\ln(-\tilde{E}_\mathrm{b})}{2
(\tilde{h}^2 + \tilde{E}_\mathrm{b})} \right.\nonumber\\[0.2cm]
& \hspace{-1.5cm} -\, \frac{\big(1 + \sqrt{1+\tilde{h}^2+
\tilde{E}_\mathrm{b}}\big)\,\big(-\tilde{E}_\mathrm{b}-2+2
\sqrt{1+\tilde{h}^2+\tilde{E}_\mathrm{b}}\big)}{4\,\sqrt{1+
\tilde{h}^2+\tilde{E}_\mathrm{b}}\, (\tilde{h}^2 +
\tilde{E}_\mathrm{b})}\,\, \ln\left( -\tilde{E}_\mathrm{b}-2+2
\sqrt{1+\tilde{h}^2+\tilde{E}_\mathrm{b}}\, \right) \nonumber
\\[0.2cm]
& \hspace{-1.5cm} \left. +\, \frac{\big(1 - \sqrt{1+\tilde{h}^2+
\tilde{E}_\mathrm{b}}\big)\,\big(-\tilde{E}_\mathrm{b}-2-2
\sqrt{1+\tilde{h}^2+\tilde{E}_\mathrm{b}}\big)}{4\,\sqrt{1+
\tilde{h}^2+\tilde{E}_\mathrm{b}}\, (\tilde{h}^2 +
\tilde{E}_\mathrm{b})}\,\, \ln\left( -\tilde{E}_\mathrm{b}-2-2
\sqrt{1+\tilde{h}^2+\tilde{E}_\mathrm{b}} \right)
\right\} \, .
\end{align}
Recognizing that the third term between curly brackets on the
r.h.s.\ of Eq.~(\ref{eq:FP0intermed}) vanishes in the limit
$\tilde{E}_\mathrm{b}\to -2 |\tilde{h}|$ with $|\tilde{h}|>1$,
we obtain
\begin{equation}
F_\mathbf{0}(-2\,\tilde{h},\tilde{h}) = \frac{\ln(2|\tilde{h}|)
- \ln[4(|\tilde{h}|-1)]}{|\tilde{h}| - 2}\, \equiv
\frac{-1}{|\tilde{h}|-2}\, \ln\left( 2\,
\frac{|\tilde{h}|-1}{|\tilde{h}|}\right) \,\, .
\end{equation}
Using this, Eq.~(\ref{eq:boundary_implicit}) with
$\tilde{E}_\mathrm{b}\to -2|\tilde{h}|$ becomes 
\begin{equation}
\gamma + \ln\left( \frac12\, \sqrt{\frac{2| \tilde{h}
|}{\tilde{\epsilon}_0}} \right) \equiv \frac12\, \ln \left(
\frac{e^{2\gamma} |\tilde{h}|}{2 \tilde{\epsilon}_0}\right)
 = \ln\left( 2\,\frac{|\tilde{h}|-1}{|\tilde{h}|}
\right)^\frac{1}{|\tilde{h}|-2} \,\, ,
\end{equation}
which can be straightforwardly solved for $\tilde{\epsilon}_0$
to yield Eq.~(\ref{eq:phaseboundary}).

In a similar fashion, we determine the boundary line dividing
regions in the $\tilde{\epsilon}_0$-$\tilde{h}$ parameter space
where $\tilde{E}_\mathrm{b}\le -1-\tilde{h}^2$ and
$-1-\tilde{h}^2 < \tilde{E}_\mathrm{b} < -2|\tilde{h}|$,
i.e., the curve where $ \tilde{E}_\mathrm{b} = -1-\tilde{h}^2$
for $|\tilde{h}|>1$. Introducing $\tilde{\delta}_\mathrm{b}
\equiv -1-\tilde{h}^2-\tilde{E}_\mathrm{b}$, we find
\begin{align}
F_0(-1-\tilde{h}^2,\tilde{h}) &= \nonumber \\ & \hspace{-2cm}
\frac{1}{4}\Bigg\{2\, \left( \tilde{h}^2-1 \right)
\underbrace{\lim_{\tilde{\delta}_\mathrm{b}\rightarrow 0}
\frac{1}{\sqrt{\tilde{\delta}_\mathrm{b}}}\left[\frac{\pi}{2}
+\arctan\left(\frac{1-\tilde{h}^2}{2
\sqrt{\tilde{\delta}_\mathrm{b}}} \right)
\right]}_{\frac{2}{\tilde{h}^2-1}} \, -\, \left( 1 + \tilde{h}^2
\right) \left[ 2 \ln\left(1+\tilde{h}^2\right) - \ln \left( 1 -
\tilde{h}^2\right)^2 \right] \Bigg\}\nonumber\\
&= 1 - \frac{1 + \tilde{h}^2}{2}\, \left[ \ln\left( 1 +
\tilde{h}^2 \right) - \ln\left| \tilde{h}^2 - 1 \right|\right]
\quad .
\end{align}
With this (using also $|\tilde{h}^2-1|\equiv \tilde{h}^2-1$ with
our assumptions), Eq.~(\ref{eq:boundary_implicit}) for
$\tilde{E}_\mathrm{b} = -1-\tilde{h}^2$ becomes 
\begin{equation}
\gamma + \ln\left( \frac12\, \sqrt{\frac{1 +
\tilde{h}^2}{\tilde{\epsilon}_0}} \right) = - 1 - \frac{1 +
\tilde{h}^2}{2}\,\, \ln\left(\frac{\tilde{h}^2-1}{\tilde{h}^2+1}
\right) \,\, ,
\end{equation}
yielding Eq.~(\ref{eq:twoRegBound}).

\section{Analytical results for fractional weights of total-spin
eigenstates in the bound state for $\mathbf{P=0}$ and
$\mathbf{\tilde{E}_b\le -1-\tilde{h}^2}$}
\label{app:spin_populations}

To calculate the fractional weights $N_{S M}$ of total-spin
eigenstates in the two-particle bound states according to
Eq.~(\ref{eq:spin_populations_integral}), integrals over the
squared magnitude of Green's-function matrix elements are
needed. For the case of vanishing COM momentum and
bound-state energy satisfying $\tilde{E}_\mathrm{b}\le -1
-\tilde{h}^2$, we can provide analytical results for the latter:
\begin{align}
&\int d^2 p\,\, |\langle 0\, 0|\hat{G}_\mathbf{0}(E_\mathrm{b},
\mathbf{p})|0\, 0\rangle |^2 = \frac{\tilde{h}^2
(\tilde{E}_\mathrm{b}-2)+\tilde{E}_\mathrm{b}^2}{2\left(
\tilde{h}^2+\tilde{E}_\mathrm{b}\right)^2 \left( -1-\tilde{h}^2
-\tilde{E}_\mathrm{b}\right)} - \frac{\tilde{h}^4 \left(\ln
\left(\tilde{E}_\mathrm{b}^2-4 \tilde{h}^2\right)-2 \ln (-
\tilde{E}_\mathrm{b})\right)}{\left(\tilde{h}^2+
\tilde{E}_\mathrm{b}\right)^3} \nonumber \\
&-\frac{\tilde{h}^4}{\tilde{E}_\mathrm{b} \left(\tilde{h}^2+
\tilde{E}_\mathrm{b}\right)^2}-\frac{\left(6 \tilde{h}^2
\tilde{E}_\mathrm{b}^3+4\tilde{h}^6(\tilde{E}_\mathrm{b}-3)+
\tilde{h}^4 (3\tilde{E}_\mathrm{b} (3 \tilde{E}_\mathrm{b}-4)
-8)+\tilde{E}_\mathrm{b}^4\right) \left(\arctan\left(
\frac{\tilde{E}_\mathrm{b}+2}{2\sqrt{-1-\tilde{h}^2
-\tilde{E}_\mathrm{b}}}\right)+\frac{\pi}{2}
\right)}{4\left( -1-\tilde{h}^2-\tilde{E}_\mathrm{b}
\right)^{3/2}\left(\tilde{h}^2+\tilde{E}_\mathrm{b}\right)^3}
\nonumber \\
\end{align}
and 
\begin{align}\label{eq:triplet11_population}
&\int d^2p\,\, |\langle 1\, \pm\! 1|\hat{G}_\mathbf{0}
(E_\mathrm{b},\mathbf{p})|0\, 0\rangle |^2 = \frac{2 ((\tilde{h}
\mp 1)\tilde{h}+1)\tilde{h}^2+(3 \tilde{h}\mp 2)\tilde{h}
\tilde{E}_\mathrm{b}+\tilde{E}_\mathrm{b}^2}{4\left(\tilde{h}^2
+\tilde{E}_\mathrm{b}\right)^2\left( -1-\tilde{h}^2
-\tilde{E}_\mathrm{b}\right)} -
\frac{\tilde{h}^2}{2\left(\tilde{h}^2+\tilde{E}_\mathrm{b}
\right)^2} + \left( \pm 4 \tilde{h}^7 \right. \nonumber\\
&+\left. \tilde{h}^4 \left(9 \tilde{E}_\mathrm{b}^2+6
\tilde{E}_\mathrm{b}-4\right)+4\tilde{h}^6 (\tilde{E}_\mathrm{b}
-1)\pm 6 \tilde{h}^5\tilde{E}_\mathrm{b}\mp 4 \tilde{h}^3
\tilde{E}_\mathrm{b}+2 \tilde{h}^2 \tilde{E}_\mathrm{b} (3
\tilde{E}_\mathrm{b} (\tilde{E}_\mathrm{b}+2)+2)\mp 2 \tilde{h}
\tilde{E}_\mathrm{b}^2(\tilde{E}_\mathrm{b}+2)\right. \nonumber
\\ & \left. + \tilde{E}_\mathrm{b}^3 (\tilde{E}_\mathrm{b}+2)
\right)\frac{\arctan\left(\frac{\tilde{E}_\mathrm{b}+2}{2
\sqrt{-1-\tilde{h}^2-\tilde{E}_\mathrm{b}}}\right)+
\frac{\pi }{2}}{8\left( -1-\tilde{h}^2-\tilde{E}_\mathrm{b}
\right)^{3/2} \left(\tilde{h}^2+\tilde{E}_\mathrm{b}\right)^3}
-\frac{\tilde{h}\left(\tilde{h}^3-(1 \mp\tilde{h})\tilde{h}
\tilde{E}_\mathrm{b}\pm \tilde{E}_\mathrm{b}^2 \right) \left(2
\ln (-\tilde{E}_\mathrm{b})-\ln \left(\tilde{E}_\mathrm{b}^2-4
\tilde{h}^2\right)\right)}{4 \left(\tilde{h}^2+
\tilde{E}_\mathrm{b}\right)^3} \,\, .
\end{align}
Note that, because $\langle 1\, 0|\hat{G}_\mathbf{0}
(E_\mathrm{b}, \mathbf p)|0\, 0\rangle=0$, $N_{1 0}=0$ for the
case $\mathbf{P} = \mathbf{0}$.

For the case of vanishing Zeeman splitting, the results above
simplify considerably, leading to
\begin{align}\label{eq:zeroH1}
&\int d^2 p\,\, |\langle 0\, 0|\hat{G}_\mathbf{0}(E_\mathrm{b},
\mathbf{p})|0\, 0\rangle |^2 \rightarrow
\frac{1}{2(-1-\tilde{E}_\mathrm{b})} -\frac{\tilde{E}_\mathrm{b}
\left(\arctan\left(\frac{\tilde{E}_\mathrm{b}+2}{2
\sqrt{-1-\tilde{E}_\mathrm{b}}}\right)+\frac{\pi}{2}\right)}{4
(-1-\tilde{E}_\mathrm{b})^{3/2}} \,\, ,  \\ \label{eq:zeroH2}
&\int d^2p\,\, |\langle 1\, \pm\! 1|\hat{G}_\mathbf{0}
(E_\mathrm{b},\mathbf{p})|0\, 0\rangle |^2 \rightarrow
\frac{1}{4(-1-\tilde{E}_\mathrm{b})}+\frac{(\tilde{E}_\mathrm{b}
+2)\left(\arctan\left(\frac{\tilde{E}_\mathrm{b}+2}{2
\sqrt{-1-\tilde{E}_\mathrm{b}}}\right)+\frac{\pi}{2}\right)}{8
(-1-\tilde{E}_\mathrm{b})^{3/2}} \,\, .
\end{align}
Using $\tilde{E}_\mathrm{b}=-1.55$ [consistent with
$\tilde{\epsilon}_\mathrm{b}=0.55$ obtained for
$\tilde{\epsilon}_0=0.1$, $\tilde{h}=0$ and $\tilde{P}=0$; see
Fig.~\ref{fig:bound_state_finite_P}(b)] in the expressions
(\ref{eq:zeroH1}) and (\ref{eq:zeroH2}), one derives $N_{00}=
0.65$ and $N_{1\,\pm 1}=0.17$ in agreement with the $\tilde{P}=
0$ results shown in Fig.~\ref{fig:triplet_finite_P}(b).

\end{appendix}


\nolinenumbers

\end{document}